\documentclass[twocolumn,prl,twocolumn,superscriptaddress]{revtex4}
\usepackage[latin9]{inputenc}
\usepackage{color}
\usepackage{amsmath}
\usepackage{amssymb}
\usepackage{graphicx}
\usepackage{tikz}

\usepackage[unicode=true,
 bookmarks=true,bookmarksnumbered=false,bookmarksopen=false,
 breaklinks=false,pdfborder={0 0 1},backref=false,colorlinks=true]
 {hyperref}
\hypersetup{
 linkcolor=magenta, urlcolor=blue, citecolor=blue, pdfstartview={FitH}, hyperfootnotes=false, unicode=true}

\makeatletter
\@ifundefined{textcolor}{}
{%
 \definecolor{BLACK}{gray}{0}
 \definecolor{WHITE}{gray}{1}
 \definecolor{RED}{rgb}{1,0,0}
 \definecolor{GREEN}{rgb}{0,1,0}
 \definecolor{BLUE}{rgb}{0,0,1}
 \definecolor{CYAN}{cmyk}{1,0,0,0}
 \definecolor{MAGENTA}{cmyk}{0,1,0,0}
 \definecolor{YELLOW}{cmyk}{0,0,1,0}
}

\usepackage{amsfonts}\usepackage{tabularx}\usepackage{dcolumn}\usepackage{bm}\usepackage{graphicx}\usepackage{epstopdf}

\hypersetup{urlcolor=blue}

\makeatother

\begin{document}

\title{Quantum Entanglement in Neural Network States}

\author{Dong-Ling Deng}
\affiliation{Condensed Matter Theory Center and Joint Quantum Institute, Department
of Physics, University of Maryland, College Park, MD 20742-4111, USA}
\author{Xiaopeng Li}
\affiliation{State Key Laboratory of Surface Physics, Institute of Nanoelectronics and Quantum Computing, and Department of Physics, Fudan University, Shanghai 200433, China}
\affiliation{Collaborative Innovation Center of Advanced Microstructures, Nanjing 210093, China} 
\affiliation{Condensed Matter Theory Center and Joint Quantum Institute, Department
of Physics, University of Maryland, College Park, MD 20742-4111, USA} 

\author{S. Das Sarma}
\affiliation{Condensed Matter Theory Center and Joint Quantum Institute, Department of Physics, University of Maryland, College Park, MD 20742-4111, USA}

\begin{abstract}
Machine learning, one of today's most rapidly growing interdisciplinary
fields, promises an unprecedented perspective for solving intricate
quantum many-body problems. %, in addition to of course solving many hard problems in computer science. 
Understanding the physical aspects of  the representative artificial neural-network states 
is recently becoming highly desirable in the applications of machine learning techniques to quantum many-body physics. 
In this paper, we explore the
data structures that encode the physical features in the network states by studying the quantum entanglement properties, with a focus on 
the restricted-Boltzmann-machine (RBM) architecture.
We prove that the entanglement entropy of all short-range RBM states
satisfies an area law for \textit{arbitrary} dimensions and bipartition
geometry. 
For long-range RBM states we show by using an {\it exact} construction  that such states could exhibit {\it volume-law} entanglement, implying a notable capability of RBM in representing quantum states with massive entanglement. Strikingly, the neural-network representation
for these states is remarkably \textit{efficient}, in a sense that
the number of nonzero parameters scales only \textit{linearly} with
the system size.
We further examine the entanglement properties of generic RBM states by randomly sampling the weight parameters of
the RBM. We find that their averaged entanglement entropy obeys volume-law scaling and meantime strongly deviates from the Page-entropy of the completely random pure states. We show that their entanglement spectrum has
\textit{no} universal part associated with random matrix theory and
bears a Poisson-type level statistics.  Using reinforcement learning, we demonstrate that RBM is capable of finding the ground state (with power-law entanglement) of a model Hamiltonian with long-range interaction.  In addition,
we show, through a concrete example of the one-dimensional symmetry-protected topological
cluster states, that the RBM representation may also be used as a
tool to analytically compute the entanglement spectrum.
Our results uncover the unparalleled power of artificial neural networks in 
representing quantum many-body states regardless of how much entanglement they 
possess, which paves 
a novel way 
to bridge computer science based machine learning techniques to outstanding quantum condensed matter physics problems. 
\end{abstract}
\maketitle

\section{Introduction}

Understanding the behavior of quantum many-body systems beyond the standard mean field paradigm is a central (and daunting)
task in condensed matter physics. 
One challenge lies in the exponential scaling of the Hilbert space dimension 
%Yet, quantum many-body problems
%are often so complex as to be provably intractable due to the exponential
%scaling of the Hilbert space dimension with the system size 
\cite{Verstraete2015Quantum,Gharibian2014Quantum,Osborne2012Hamiltonian}.
In principle, a complete description of a generic many-body state 
requires an exponential amount of information, rendering the problem
unattainable even numerically. Yet, fortunately physical states
 usually only access a tiny corner of the entire Hilbert
space and can often be characterized with much less classical resources. 
Constructing efficient representations of such states 
 are thus of crucial importance in tackling quantum
many-body problems. Notable examples include quantum states with area-law
entanglement \cite{Eisert2010Area}, such as ground states of local
gapped Hamiltonians \cite{Hastings2007Area} or the eigenstates of
many-body localized systems \cite{Friesdorf2015MBL}, which can be
efficiently represented in terms of matrix product states (MPS) \cite{Fannes1992Finitely,Perez2007Matrix,Schollwock2011Density}
or tensor-network states in general \cite{Verstraete2008Matrix,Gu2009Tensor,Orus2014Practical}.
These compact representations of quantum states play a vital role
and are indispensable for tackling a variety of many-body problems
ranging from the classification of topological phases \cite{chen2012symmetry,Chen2013Symmetry}
to the construction of the Ads/CFT correspondence \cite{Swingle2012Entanglement,Swingle2012Constructing}.
In addition, they are also the backbones of a number of efficient
classical algorithms for solving intricate many-body problems, e.g.,
DMRG (density-matrix renormalization group) \cite{White1992Density,Schollwock2005TheDMRG,Schollwock2011Density},
TEBD (time-evolving block decimation) \cite{Vidal2003Efficient},
PEPS \cite{Orus2014Practical,Verstraete2008Matrix} (projected entangled
pair states), and MERA (multiscale entanglement renormalization ansatz)
methods \cite{Vidal2007Classical,Vidal2008Class}. 
Recently, a novel neural-network representation of quantum many-body states has been introduced \cite{Carleo2016Solving} in solving many-body problems with machine learning techniques. However, the entanglement properties (which are crucial for the 
renowned MPS/tensor-network representations) of these neural-network states remain unknown. %In this paper, we study their entanglement properties.
In this paper, we fill this crucial gap by studying the entanglement properties of these many-body neural-network quantum states both analytically and numerically. Our work provides an important connection between the physical properties of many-body quantum entanglement and the computer science properties of neural network based machine leaning.

%However for the recently proposed neural-network states, the entanglement properties remain unknown, 
%which sets a roadblock of using machine-learning techniques in quantum many-body physics. 
%Inspired by this,
%in this paper we study the entanglement properties of quantum states
%with a neural-network representation, 
%which is essential for the machine-learning
%approach to many-body problems. 

Machine learning %, one of today's most rapidly growing fields, 
is the
core of artificial intelligence and data science \cite{Michalski2013Machine}.
It powers many aspects of modern society and its applications have
become ubiquitous throughout science, technology, and commerce \cite{Jordan2015Machine,Lecun2015Deep}. In fact, perhaps because of the dominant presence of big data in our modern world, the terms artificial intelligence, machine learning, neural networks, deep learning, etc. have generically entered the lexicon of the cultural world, well outside the technical world of computer science where they originated, often appearing in everyday press and popular articles or stories---for example, the software technology underlying automated self-driving cars depends crucially on artificial intelligence and machine learning.
Within physics, applications of machine-learning techniques have 
recently been invoked in various contexts such 
as  gravitational wave analysis
\cite{Rahul2013Application,Abbott2016Observation}, black hole
detection \cite{Pasquato2016Detecting}, material design \cite{Kalinin2015Big}, 
and classification of the classical liquid-gas transitions~\cite{Schoenholz2016Structural}. Very recently, these techniques
have been introduced to many-body quantum condensed-matter physics, raising
considerable interest across different communities \cite{Arsenault2015Machine,Zhang2016Triangular,Carrasquilla2016Machine,van2016Learning,
Wang2016Discovering,Torlai2016Learning,Broecker2016Machine,
Carleo2016Solving,Chng2016Machine,Torlai2016Neural,
Amin2016Quantum,Liu2016Self,Huang2016Accelerate,Aoki2016restricted,Biamonte2016Quantum}.
Exciting progress has been made in identifying quantum phases and transitions
among them (either conventional symmetry-broken \cite{Wang2016Discovering,Chng2016Machine,Carrasquilla2016Machine,Broecker2016Machine}
or topological phases \cite{Zhang2016Triangular}), modeling thermodynamic
observables \cite{Torlai2016Learning}, constructing decoders for
topological codes \cite{Torlai2016Neural}, accelerating Monte Carlo
simulations \cite{Liu2016Self,Huang2016Accelerate}, and establishing
connections to renormalization group techniques \cite{Beny2013Deep,Mehta2014Exact},
etc. 
%{\color{red}[Please consider rewritting the above sentence. It does not seem to flow. ---XL]}
 In addition, machine-learning ideas have also been explored in
measuring quantum entanglement and wave-function tomography through
the analyses of data extracted from quantum gas microscopes in cold atom experiments
\cite{Tubman2016Measuring}. The fledgling field of machine learning applications in physics appears to be in its rapidly growing early phase with many expected future breakthroughs as it matures.

\begin{figure}
\includegraphics[width=0.46\textwidth]{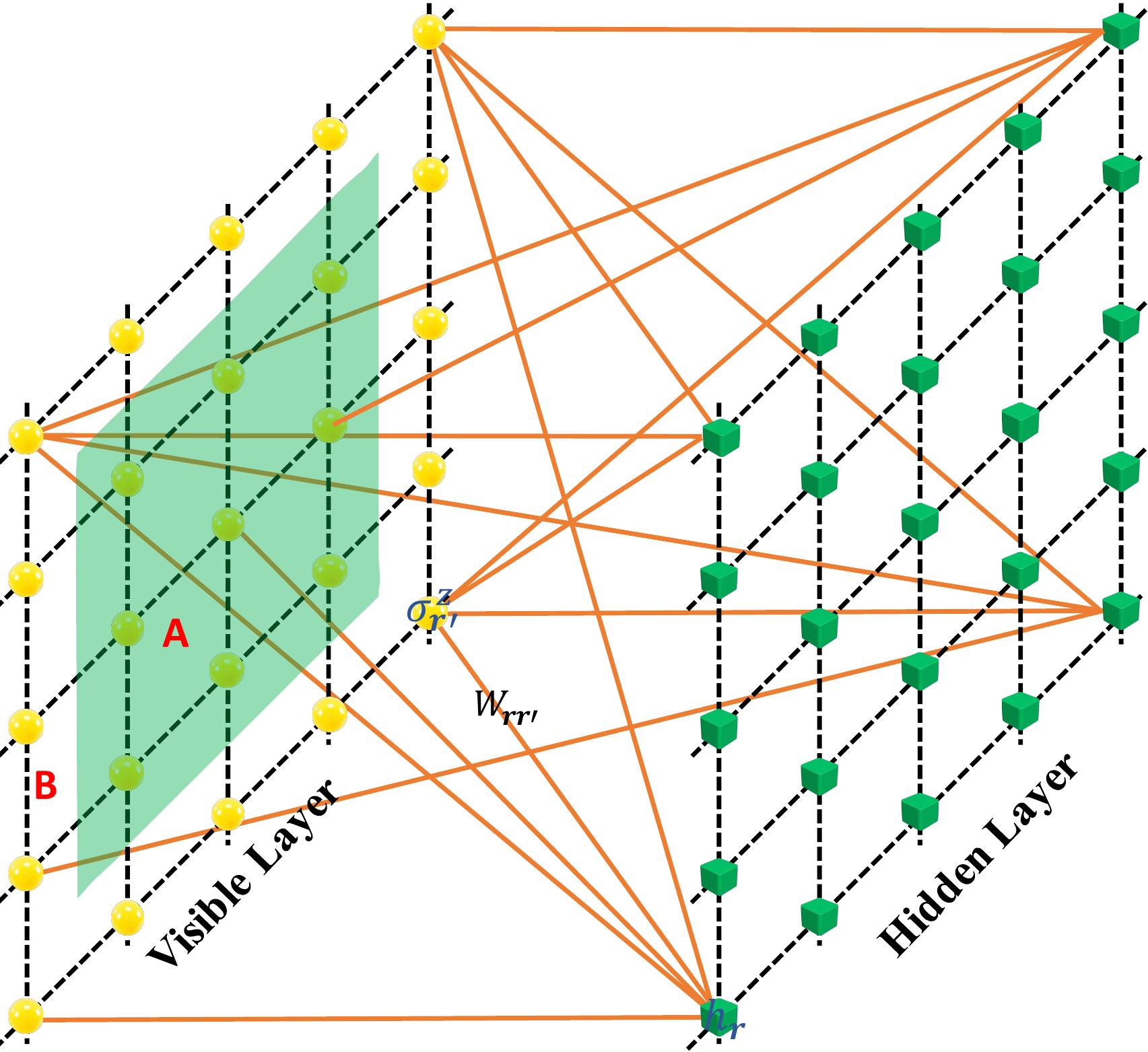}

\caption{A 2D pictorial illustration of artificial-neural-network quantum states
in the restricted-Boltzmann-machine architecture. The yellow balls
(green cubes) denote the neurons on the visible (hidden) layer, corresponding
to the physical (auxiliary) spins. The brown lines show the connections
between visible and hidden neurons, with the weight parameter denoted
by $W_{\mathbf{r}\mathbf{r}'}$ (only a small portion of the connections
is shown for best visualization). Here, we also show a typical bipartition
of the system into two subsystems $A$ and $B$ in order to study
the entanglement properties of the neural-network states. \label{fig:NNetw}}
\end{figure}

From the numerical perspective, the applications of machine-learning
techniques to many-body problems would  rely vitally on the underlying data
structures of the artificial neural networks, whose connections to the entanglement 
features of the corresponding quantum states are  particularly desirable to be addressed. 
% However, despite rapid progress as mentioned above,
%no study on the entanglement structures of neural-network quantum
%states has been reported so far to the best of our knowledge. 
In this
paper, we study the entanglement properties, such as the entanglement entropy and spectrum, of the neural-network states. We focus on the 
quantum states represented by the restricted Boltzmann
machine (RBM), which is a stochastic artificial neural network with
widespread applications \cite{Hinton2006Reducing,Salakhutdinov2007Restricted,Larochelle2008Classification,Amin2016Quantum}.
We first prove the general result that all short-range RBM states obey an entanglement area-law, 
independent of dimensionality and bipartition geometry.
Since the one-dimensional (1D) symmetry-protected topological (SPT) cluster states and toric code states (in both 2D and 3D) have an exact short-range RBM representation \cite{Deng2016Exact}, it follows immediately that they all have area-law entanglement. 
For  long-range RBM states, calculating their entanglement entropy and spectrum analytically is very challenging (if not impossible) and we thus resort to numerical simulations. We randomly sample the weight parameters of the RBM states and compute their entanglement entropy and spectrum. We find that their entanglement entropy exhibits a volume-law scaling 
in general.
%For long-range RBM states, we randomly sample their weight parameters and find that their entanglement entropy exhibits a volume-law scaling 
%in general. %for their entanglement entropy on average. 
However, surprisingly their entropy is noticeably less than the Page-entropy for random pure states,
and their entanglement spectrum has \textit{no} universal part associated
with random matrix theory and bears a Poisson-type level statistics.
This indicates that the RBM states with random weight parameters live
in a very restricted subspace of the entire Hilbert space (in spite of manifesting a volume law entanglement entropy) and are
not irreversible\textemdash namely there exists an efficient algorithm
to completely disentangle these states \cite{Chamon2014Emergent}. 

In addition, we analytically construct a family of RBM states with maximal volume-law entanglement. These states \textit{cannot} be described in terms of
matrix product states or tensor-network states with a computationally
tractable bond dimension. In sharp contrast, their RBM representation
is remarkably \textit{efficient}, requiring only a small number of
parameters that scales \textit{linearly} with the system size. This
shows, in an \textit{exact} fashion and the most explicit way, the
unparalleled power of neural networks in describing many-body quantum
states with large entanglement. Unlike MPS/tensor-network states, entanglement is not the limiting factor for the efficiency of the neural-network representation.  As an important consequence, we are able to calculate (through a reinforcement-learning scheme \cite{Sorella2007Weak,Carleo2016Solving}) the ground state, whose entanglement has a power-law scaling with system size, of a spin Hamiltonian with long-range interaction.
Finally, we show that the RBM representation
could also be used as a tool to analytically compute the entanglement
entropy and spectrum for certain quantum states with short-range RBM
descriptions. We demonstrate this by using a concrete example of the 1D
SPT cluster states. Our results not only demonstrate explicitly the exceptional power of artificial neural networks in representing quantum many-body states, but also reveal some crucial aspects of their data structures, which provide a valuable guide for the emerging new field of machine learning and many-body quantum physics.

%{\color{red} [As written, the main message(s) is very unclear. ---XL]} 

\section{Neural-network representation and quantum entanglement: concepts
and notations}

An artificial-neural-network representation of quantum many-body states
has recently been introduced by Carleo and Troyer in Ref. \cite{Carleo2016Solving},
where they demonstrated the remarkable power of a reinforcement-learning
approach in calculating the ground-state or simulating
the unitary time evolution of complex quantum systems with strong
interactions. 
We show elsewhere that this representation can be used to describe topological states, 
even for those with long-range entanglement \cite{Deng2016Exact}. To start
with, let us first briefly introduce this representation in the RBM
architectures. We consider a quantum system with $N$ spins living
on a $d$ dimensional cubic lattice 
$\Xi=(\sigma_{\mathbf{r}_{1}},\sigma_{\mathbf{r}_{2}},\cdots,\sigma_{\mathbf{r}_{N}})$. 
Correspondingly we introduce $\Xi_Y$ for spins in a subsystem $Y$ as 
$\Xi_Y =  \{ \sigma_{\bf r} : {\bf r} \in Y\} $. 
The geometric details of the lattice do not matter. Here, we choose
cubic lattices and focus on spin-$\frac{1}{2}$ (qubits) systems for
simplicity. An RBM neural network contains two layers, one visible
layer with $N$ nodes (visible neurons) corresponding to the physical
spins, the other a hidden layer with $M$ auxiliary nodes $(h_{\mathbf{r}_{1}},h_{\mathbf{r}_{2}},\cdots,h_{\mathbf{r}_{M}})$
(hidden neurons). The neurons in the hidden layer are connected to
these in the visible layer, but there is no connection among neurons
in the same layer (see Fig. \ref{fig:NNetw} for an 2D illustration).
The RBM neural-network representation of a quantum state is obtained
by tracing out the hidden neurons \cite{Carleo2016Solving} : 
\begin{eqnarray}
\Phi_{M}(\Xi;\Omega) & = & \sum_{\{h_{\mathbf{r}}\}}e^{\sum_{\mathbf{r}}a_{\mathbf{r}}\sigma_{\mathbf{r}}^{z}+\sum_{\mathbf{r}'}b_{\mathbf{r}'}h_{\mathbf{r}'}+\sum_{\mathbf{r}\mathbf{r}'}W_{\mathbf{r}'\mathbf{r}}h_{\mathbf{r'}}\sigma_{\mathbf{r}}^{z}},\;\label{eq:RBM states}
\end{eqnarray}
where $\{h_{\mathbf{r}}\}=\{-1,1\}^{M}$ denotes the possible configurations
of the hidden spin variables and the weights $\Omega=(a_{\mathbf{r}},b_{\mathbf{r}'},W_{\mathbf{r}\mathbf{r}'})$
are parameters needed to be trained to best represent the many-body
quantum state. It is worthwhile to mention that the RBM state defined
in Eq. (\ref{fig:NNetw}) is a variational state with its amplitude
and phase specified by $\Phi_{M}(\Xi;\Omega)$. The actual quantum
state should be understood as (up to an irrelevant normalization constant)
$|\Psi(\Omega)\rangle\equiv\sum_{\Xi}\Phi_{M}(\Xi;\Omega)|\Xi\rangle$,
similar to the Laughlin-like description of the resonating-valence-bond
ground state of the exactly-solvable Haldane-Shastry model \cite{Haldane1988Exact,Shastry1988Exact}. 

We remark that RBMs can be trained in either supervised or unsupervised
ways, and in the machine-learning community RBMs have had successful
applications in classification \cite{Larochelle2008Classification},
dimensionality reduction \cite{Hinton2006Reducing}, feature learning
\cite{Coates2011Analysis}, and collaborative filtering \cite{Salakhutdinov2007Restricted},
etc. Mathematically, the ability of the RBM to approximate any many-body
state is assured by representability theorems \cite{Kolmogorov1963Representation,Le2008Representational,Hornik1991Approximation}.
Nevertheless, the approximation may require a huge number (exponential
in system size) of neurons and parameters, thus rendering the representation
impractical, especially in numerical simulations. A question
of both theoretical and practical interest is: what kind of many-body
quantum states can be efficiently described by RBMs with a numerically feasible
number of neurons and parameters? It is now established that entanglement
plays a crucial role in determining whether a quantum state can be
efficiently represented by MPS/tensor-network or not. Quantum states with volume-law
entanglement cannot be described efficiently by MPS/tensor-network and thus cannot
be simulated efficiently by DMRG, PEPS, or MERA. In sharp contrast,
as we will show in the following sections, RBMs are indeed capable
of efficiently describing  certain specific quantum states with volume-law entanglement,
giving rise to the great potential of numerically simulating these
states with new machine learning algorithms based on RBMs.

In this paper, we study the quantum entanglement properties of the
RBM states. In particular, we investigate the entanglement entropy,
the R\'{e}nyi entropy \cite{Amico2008Entanglement}, and the entanglement
spectrum \cite{Li2008Entanglement}, which are three of the most broadly
used quantities for characterizing many-body entanglement of a pure quantum
state. These quantities can be defined as follows: considering a pure
many-body quantum state $|\psi\rangle$, we divide the system into
two subregions, $A$ and $B$ (a typical bipartition of a 2D system
is shown in Fig.\ref{fig:NNetw}) . We then construct the reduced
density matrix of subsystem $A$ by tracing out the degree of freedom
in $B$: $\rho_{A}(|\psi\rangle)=\text{Tr}_{B}(|\psi\rangle\langle\psi|)$.
The $\alpha$-th order R\'{e}nyi entropy is defined as 
\begin{eqnarray*}
S_{\alpha}^{A} & \equiv & \frac{1}{1-\alpha}\log[\text{Tr}(\rho_{A}^{\alpha})].
\end{eqnarray*}
The zeroth order ($\alpha=0$) R\'{e}nyi entropy is related to the rank,
namely, the number of nonzero singular values of $\rho_{A}$. When
$\alpha\rightarrow1$, the first order R\'{e}nyi entropy reduces to the
von Neumann entropy,
\begin{eqnarray*}
S_{1}^{A} & \equiv & -\text{Tr}(\rho_{A}\log\rho_{A}).
\end{eqnarray*}
In the literature, the entanglement entropy usually means the von
Neumann entropy. However, throughout this paper we do not differentiate
between the entanglement entropy and the R\'{e}nyi entropy, since most of the
results are valid for the R\'{e}nyi entropy to all orders. To define the
entanglement spectrum, we first define the entanglement Hamiltonian
by taking the log of $\rho_{A}$: 
\begin{eqnarray*}
H_{\text{ent}} & \equiv & -\log\rho_{A},
\end{eqnarray*}
and then the entanglement spectrum is defined as the spectrum of $H_{\text{ent}}$.
We mention that entanglement is nowadays a central concept in many
branches of quantum physics. In condensed matter physics, entanglement
entropy and spectrum have proven to be powerful tools characterizing
topological phases \cite{Eisert2010Area,Levin2006Detecting,Kitaev2006Topological,
Li2008Entanglement,Biddle2011Entanglement},
quantum phases transitions \cite{Osterloh2002Scaling,Osborne2002Entanglement,Gu2004Entanglement},
and many-body localization \cite{Pal2010Manybody,Bauer2013Area,Nandkishore2015many},
etc. A number of theoretical proposals have been introduced to measure
entanglement entropy \cite{Alves2004Multipartite,Daley2012Measuring,Abanin2012Measuring,Hauke2016Measuring}
and spectrum \cite{Pichler2016Measurement} in many-body systems.
Notably, experimental measurements of the second-order R\'{e}nyi entropy have
been achieved in recent cold-atom experiments in optical lattices
\cite{Islam2015Measuring,Kaufman2016Quantum}. We expect our study
of entanglement properties of neural-network states would also provide
novel inspiration in this context. 

\section{area-law entanglement for short-range neural-network states}
\label{sec:AreaLaw} 

We start with short-range RBM states and prove that they obey an area-law entanglement scaling, namely the amount of entanglement between a subsystem
and its complement scales at most as the surface area or the boundary
rather than the volume of the subsystem \cite{Eisert2010Area}. Historically,
the study of entanglement area laws is inspired by the holographic
principle in black hole physics, where the Bekenstein-Hawking entropy
of a black hole is believed to scale as its boundary surface rather
than its volume \cite{Hawking2001Desitter}. It has been argued that
the origin of the black hole entropy is the quantum entanglement between
the inside and the outside of the black hole \cite{Bekenstein1973Black,Bombelli1986Quantum,Srednicki1993Entropy}.
Although it is apparent that entanglement in \textquotedblleft natural\textquotedblright{}
quantum systems should roughly live on the boundary and many numerical
simulations indeed support this intuition, rigorously proving the area law
for a given family of quantum states is notoriously challenging and
often involves sophisticated mathematical techniques \cite{Eisert2010Area}.
A breakthrough was first made by Hastings in Ref. \cite{Hastings2007Area},
where he proved an entanglement area law for the ground states of
1D gapped local Hamiltonians by using the Lieb-Robinson bound \cite{Lieb1972Finite}.
More recently, this proof has been simplified and generalized to ground
states with a finite number of degeneracy by a combinatorial approach
based on Chebyshev polynomials \cite{Arad2013Area,Huang2014Area}.
Unfortunately, both the Lieb-Robinson bound approach and the combinatorial
approach seem unlikely to carry over to the case of higher dimensions.
Establishing the area-law entanglement for ground states of gapped Hamiltonians in more than
one dimension remains a major open problem (and arguably the most
important one) in the field of Hamiltonian complexity \cite{Osborne2012Hamiltonian}.

Here we prove that short-range RBM states obey the area law of entanglement
in \textit{any} dimension for \textit{arbitrary} bipartition geometry.
To be precise, we call a RBM state a $\mathcal{R}$-range RBM state
if each hidden neuron is only connected to these visible neurons within
a $\mathcal{R}$ neighborhood, i.e., $W_{\mathbf{r}\mathbf{r}'}=0$
if $|\mathbf{r}-\mathbf{r}'|>\mathcal{R}$. 
%{\color{red}[It seems we need to change quite a bit if we switch to local connectivity, so maybe we should decide first.]} 
For instance, in Ref.
\cite{Deng2016Exact} we have demonstrated that both the 1D SPT cluster
states and toric code states (both 2D and 3D) can be represented {\it exactly} by RBMs
with hidden neurons being connected only to nearest visible neurons.
These states are 1-range RBM states. For general $\mathcal{R}$-range
RBM states, we have the following theorem:

\begin{figure}
\includegraphics[width=0.4\textwidth]{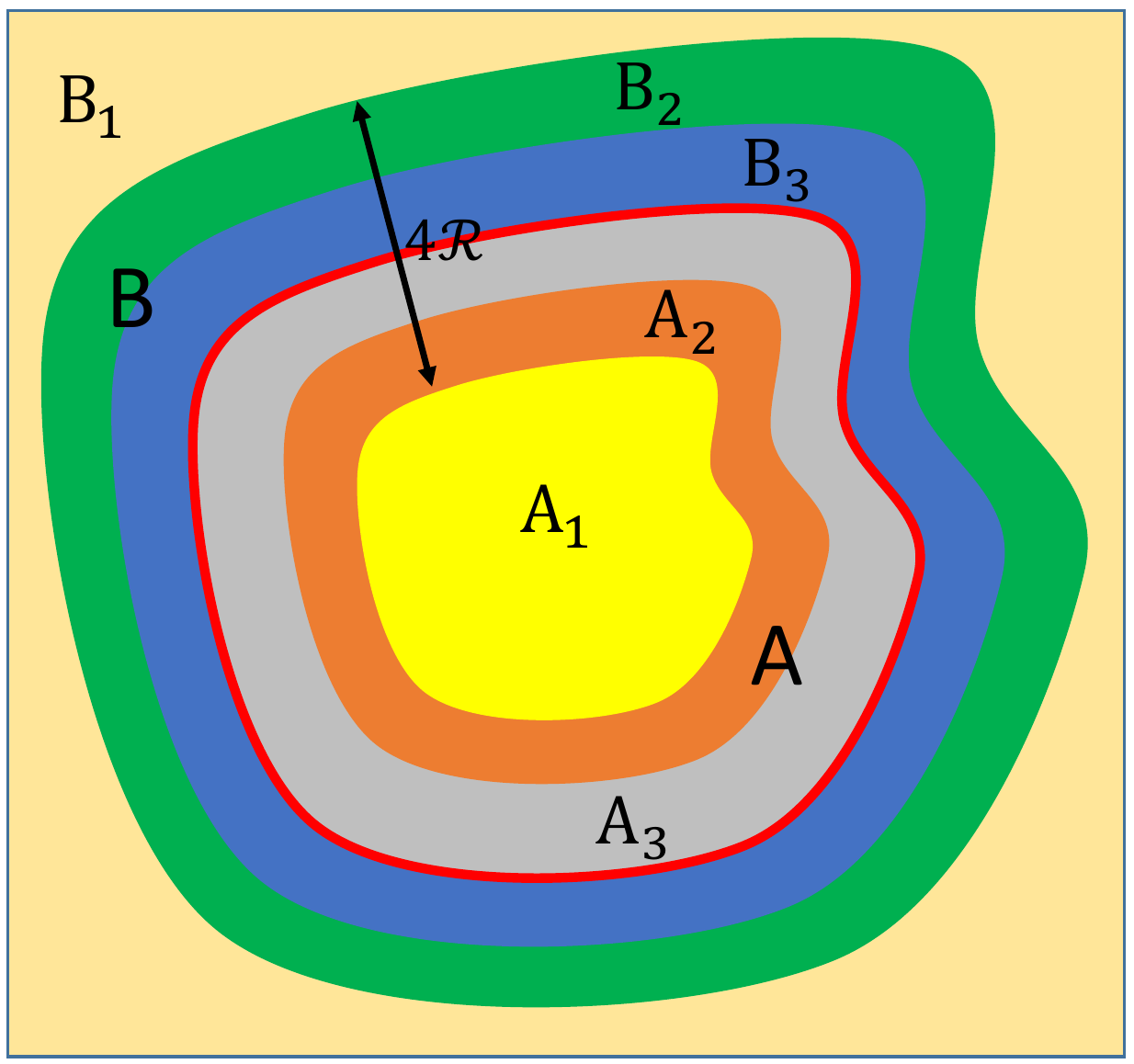}

\caption{A sketch for the proof of area-law entanglement for short-range restricted-Boltzmann-machine
states. The system is divided into two subsystems $A$ and $B$ with
the red line showing the interface boundary. In order to show that
the R\'{e}nyi entropy $S_{\alpha}^{A}$ obey an area law for any $\alpha$,
the subsystem $A$ ($B$) is further divided into three parts $A_{1}$,
$A_{2}$ and $A_{3}$ ($B_{1}$, $B_{2}$ and $B_{3}$). Since the
neural-network is short-range, the $\Gamma_{\mathbf{r}}$-factors
with $\mathbf{r}\in A_{1}\cup A_{2}$ ($\mathbf{r}\in B_{1}\cup B_{2}$
) is independent of spin configurations in region $B$ (A), thus we
can group the spins in region $A_{1}$ and $B_{1}$ with their $\Gamma_{\mathbf{r}}$-factors.
The entropy of the reduced density matrix $\rho_{A}$ then only depends
on the degree of freedom in region $A_{2}\cup A_{3}$, which is proportional
to the surface area of region $A$. This gives us a clear geometric
picture of why $S_{\alpha}^{A}$ is upper bounded by the surface area
of $A$, up to an unimportant scaling constant as given in Eq.(\ref{eq:AreaLaw}).
\label{fig:Sketch-AreaLaw} }
\end{figure}

\textit{Theorem 1.} \textemdash For a $\mathcal{R}$-range RBM state,
the R\'{e}nyi entropy for all  orders satisfies
\begin{eqnarray}
S_{\alpha}^{A} & \leq & 2\mathcal{S}(A)\mathcal{R}\log2,\quad\forall\alpha,\label{eq:AreaLaw}
\end{eqnarray}
where $\mathcal{S}(A)$ denotes the surface area of the subsystem
$A$. This area law is valid in any dimension and for arbitrary bipartition
geometry. 

\textit{Proof. }For RBMs, since there is no intra-layer connections
between neurons, we can explicitly factor out the hidden variables
and rewrite $\Phi_{M}(\Xi;\Omega)$ in a product form: 
$\Phi_{M}(\Xi;\Omega)=\prod_{\mathbf{r}}e^{a_{\mathbf{r}}\sigma_{\mathbf{r}}^{z}}\prod_{\mathbf{r}'}
\Gamma_{\mathbf{r'}}(\Xi_{\underline{{\bf r}'}})$, 
where we have introduced a local subregion notation 
$\underline{{\bf r}_0} \equiv \{ {\bf r }: |{\bf r} - {\bf r}_0 | < R \} $, and 
a local function $\Gamma_{\mathbf{r'}}(\Xi_{\underline{ {\bf r}'}})=2\cosh(b_{\mathbf{r'}}+\sum_{\mathbf{r}}W_{\mathbf{r'}\mathbf{r}}\sigma_{\mathbf{r}}^{z})$.
We call $\Gamma_{\mathbf{r'}}(\Xi _{\underline{ {\bf r}'}} )$ the $\Gamma_{\mathbf{r}'}$-factor
for the hidden neuron at $\mathbf{r}'$. Moreover, by the definition
of $\mathcal{R}$-range RBM, the values of $\Gamma_{\mathbf{r}'}$-factors
only depend on the configuration of these visible neurons (physical
spins) within a $\mathcal{R}$ neighborhood (denoted by $\mathcal{N}_{\mathbf{r}'}(\mathcal{R})$).
We can thus simplify $\Gamma_{\mathbf{r'}}(\Xi_{\underline{ {\bf r}'}})$ as $\Gamma_{\mathbf{r'}}(\Xi_{\underline{ {\bf r}'}})=2\cosh(b_{\mathbf{r}'}+\sum_{\mathbf{r}\in\mathcal{N}_{\mathbf{r}'}(\mathcal{R})}W_{\mathbf{r'}\mathbf{r}}\sigma_{\mathbf{r}}^{z})$.
This indicates the locality feature of $\Gamma_{\mathbf{r}'}$-factors,
which is the origin of the area law. 

In order to utilize the locality-feature of $\Gamma_{\mathbf{r}'}$-factors,
we can further divide the subregion $A$ ($B$) into three parts,
$A_{1}$, $A_{2}$, and $A_{3}$ ($B_{1}$, $B_{2}$, and $B_{3}$),
as illustrated in Fig. \ref{fig:Sketch-AreaLaw}. 
Explicitly,  the subregion with ${\bf r}$ directly coupled (via one hidden neuron) to 
the lattice sites in $B$ is defined to be $A_3$, the one directly coupled to $A_3$ within $A$ is defined to  be $A_2$, and 
the rest of $A$ is $A_1$. The subregions $B_{1,2,3}$ are introduced correspondingly. 
One can regard the
subregions $A_{2}$, $A_{3}$, $B_{2}$, and $B_{3}$ as hypersurfaces
with thickness $\mathcal{R}$ in high dimensions. 
%For a given region $X$, we denote the spin configuration within this region as $\Xi_{X}$.
Let $Y=A_{2}\cup A_{3}\cup B_{2}\cup B_{3}$, we can rewrite the $\mathcal{R}$-range
RBM state as 
\begin{eqnarray}
|\Psi(\Omega)\rangle & = & 
\sum_{\Xi_{Y}}\prod_{\mathbf{r}'\in A_{3}\cup B_{3}} 
	\Gamma_{\mathbf{r}'}(\Xi_{\underline{ {\bf r}'}})|\varphi_{A}\rangle|\varphi_{B}\rangle,\label{eq:PsiABparts}
\end{eqnarray}
where $|\varphi_{A}\rangle=\sum_{\Xi_{A_{1}}}\prod_{\mathbf{r}'\in A_{1}\cup A_{2}}
	\Gamma_{\mathbf{r}'}(\Xi_{\underline{ {\bf r}'}})|\Xi_{A}\rangle$
and $|\varphi_{B}\rangle=\sum_{\Xi_{B_{1}}}\prod_{\mathbf{r}'\in B_{1}\cup B_{2}}\Gamma_{\mathbf{r}'}(\Xi_{\underline{ {\bf r}'}})|\Xi_{B}\rangle$.
From Eq. (\ref{eq:PsiABparts}), we only have at most $2^{|Y|}$ terms,
with $|Y|$ denoting the number of spins in region $Y$, in the summation
and each term is a tensor product of orthogonal states of $A$ and
$B$. This gives the upper bound in Eq. (\ref{eq:AreaLaw}) after
tracing out the degrees of freedom in region $B$. We stress two crucial
aspects of Eq. (\ref{eq:PsiABparts}): (i) $|\varphi_{A}\rangle$
is independent of spin configurations in region $B$ and $|\varphi_{B}\rangle$
is independent of spin configurations in region $A$; (ii) the coefficients
$\prod_{\mathbf{r}'\in A_{3}\cup B_{3}}\Gamma_{\mathbf{r}'}(\Xi_{\underline{ {\bf r}'}})$
for each orthogonal components $|\varphi_{A}\rangle|\varphi_{B}\rangle$
are independent of spin configurations outside $Y$. (i) and (ii)
are crucial for the validity of the proof and they are made
evident by the deliberate partitions of both subregions $A$ and $B$
further into three smaller parts. 

We emphasize that in the above proof, we did not specify the dimensionality
or the geometry of the bipartition. The proof works for any dimension
and any bipartition of the system. 
%This is different from the
%notable area-law proofs of the ground states of local Hamiltonians
%with a finite gap \cite{Hastings2007Area,Arad2013Area,Huang2014Area},
%or the quantum states with exponential decay of correlations \cite{Brandao2013Area},
%where the discussions are specifically restricted to 1D and the methods used do
%not seem easily generalizable to higher dimensions. 
%{\color{red} [This is not ``comparing apple to apple". I would suggest to remove (or soften) the above sentence. ---XL ]} 
Thus, it might shed new light on the important challenging problem of proving the entanglement area law for local gapped Hamiltonians in higher dimensions \cite{Hastings2007Area,Arad2013Area,Huang2014Area,Brandao2013Area}, given the possibility that all ground states of these
Hamiltonians perhaps are representable by short-range RBMs, although a
rigorous proof of this still remains unclear.
%Our proof sheds new light on this important challenging problem and may provide a promising
%approach, given the possibility that all ground states of local gapped
%Hamiltonians perhaps be representable by short-range RBMs, although a
%rigorous proof of this still remains unclear. 
Intuitively, one can increase the number of hidden neurons to increase the number of local weight parameters. When there are enough free parameters, the corresponding RBMs should be able to represent the ground states of general local gapped Hamiltonians.   This would work because of a crucial aspect of our proof---the numbers of hidden neurons and weight parameters are unlimited as long as the connections are finite-ranged.

As shown in our previous work \cite{Deng2016Exact}, the 1D SPT cluster
states, the toric code states in both 2D and 3D,  and the low-energy excited states with abelian anyons of the toric code Hamiltonians can all be represented exactly by
short-range RBMs with $\mathcal{R}=1$. An immediate corollary of
the above theorem is that the entanglement of all these states fulfill
an area law. In fact, based on the RBM representation, one can even
compute analytically the entanglement spectrum of the
1D SPT cluster state, as we will show in  a latter section (see Sec.~\ref{sec:EntSpectrum}). It
is important to clarify that although the RBMs are short-range, their
represented quantum states can capture long-range entanglement. The
RBM representation of the toric code states (both in 2D and 3D), which have  intrinsic
topological orders (long-range entangled), are such examples. The
area law of short-range RBM states does \textit{not} imply short-range
entanglement. This distinction between short-ranged in the RBM sense and short-ranged in the entanglement sense is an important point. % or exponential decay of correlations. 

In 1D, the area-law bound in Eq. (\ref{eq:AreaLaw}) gives rise to
an interesting relation between the RBM and MPS representations of
quantum many-body states. It has been proved that a bounded R\'{e}nyi
entropy of all the orders in 1D necessarily guarantees an efficient
MPS representation \cite{Verstraete2006Matrix} (note that counterexample
does exist if only the von Neumann entropy is bounded \cite{Schuch2008Entropy}).
As a result, our area-law results imply that all 1D short-range RBM
states can be efficiently described in terms of MPS. However, the
validity of the inverse statement is unknown. It would
be interesting to find out whether all MPS descriptions with small bond dimensions
have efficient RBM representations or not, and if so, what the
general procedure is for recasting MPS into RBM states. It would also be
 interesting to investigate the relations between higher dimensional
RBM states and tensor-network states, PEPS, or MERA. 
Nonetheless it is worth emphasizing here that the entanglement scaling 
of RBM states is sharply distinctive from MPS---the maximal entanglement entropy 
of  a ${\cal R}$-range RBM state scales linearly with ${\cal R}$ whereas a bond-dimension ($\chi$) MPS 
has an entanglement entropy scaling as $\log \chi$. This implies that even a RBM state can be generically converted to a MPS, the parameterization in RBM states is much more efficient for representing highly entangled quantum states. 

We remark that our rigorous proof of entanglement area law for short-range RBMs  provides a valuable guide for some practical numerical calculations. For instance, in some circumstances we know that the problem may only involve a small amount (an area law) entanglement, then we may use short-range, rather than long-range, RBMs to reduce the number of parameters and consequently speed up the calculations (we have tested this in a numerical experiment of finding the ground state of the transverse-field Ising model via reinforcement learning and a considerable speedup has indeed been obtained). On the other hand, if the problem to be solved involves large entanglement (such as some quantum criticality or quantum dynamic problems), then  short-range RBMs will necessarily not work and we should choose a long-range RBM to begin with.

\section{Volume-law entanglement in long-range neural-network states}

In the last section, we proved that all short-range RBM states satisfy
an area-law entanglement. What about RBM states with long-range neural
connections? From the linear-in-${\cal R}$ entanglement-entropy scaling of ${\cal R}$-range RBM states derived in the last section, we would anticipate that long-range RBM states could exhibit volume-law entanglement. In this section, we explicitly show that this is indeed true by a rigorous exact construction and a numeric benchmark. 

\begin{figure}
\includegraphics[width=0.49\textwidth]{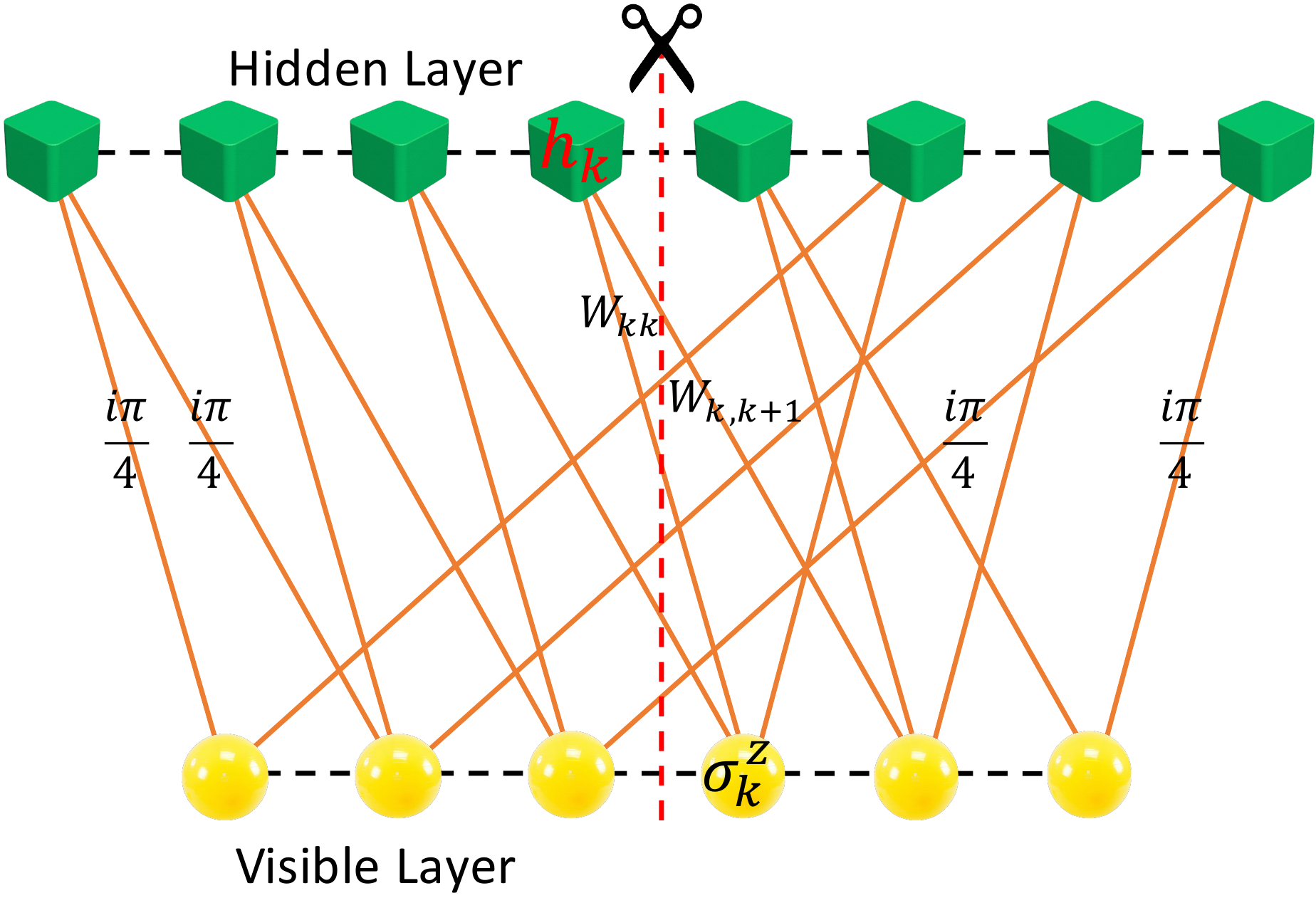}

\caption{An illustration of the constructed $1D$ neural-network state with
a maximal volume-law entanglement entropy. This restricted Boltzmann
machine has $N$ visible and $\lfloor\frac{3N}{2}\rfloor-1$ hidden
neurons. For $1\leq k\leq N-1$ ($N\leq k\leq\lfloor\frac{3N}{2}\rfloor-1$)
, The $k$-th hidden neuron is connected to two visible neurons at
sites $k$ and $k+1$ ($k+1-N$ and $k+1-\lceil\frac{N}{2}\rceil$)
with connection weight parameters equal $\frac{i\pi}{4}$. The on-site
potentials for the visible neurons are chosen to be zero $a_{k}=0$
($\forall k$). For the hidden neurons, $b_{k}$ is chosen as: $b_{k}=-\frac{i\pi}{4}$
($1\leq k\leq N-1$) and $b_{k}=\frac{\pi i}{2}$ ($N\leq k\leq\lfloor\frac{3N}{2}\rfloor-1$).
The scissors show a cut of the system into two subsystems ($A$ and
$B$) with equal sizes and for this bipartition the R\'{e}nyi entropy
is $S_{\alpha}^{A}=\lfloor\frac{N}{2}\rfloor\log2$, proportional
to the system size. This is also the maximal amount of entropy one
can have for a system with $N$ qubits. 
%{\color{red} [Maybe it is better to label $1$ to $N-1$, $\ldots$ in the figure directly. ---XL ]} 
\label{fig:VolumeLaw1D}}
\end{figure}

\subsection{Exact construction of maximal volume-law entangled neural-network states} 

%In the above section, we numerically studied the entanglement properties of the long-range RBM states with random parameters and find that their entanglement entropy exhibits a volume law on average. 
Here we construct analytically families of neural-network states
with volume-law entanglement. These states are \textit{exact }and
have unified \textit{closed-form} RBM representations. More strikingly,
the RBM representation of these states is surprisingly efficient\textemdash the
number of nonzero parameters scales only \textit{linearly} with the
system size! We stress that efficient representations of quantum states
play a vital role in solving many-body problems, especially when numerical
approaches are employed. A prominent example is the advantageous usage
of MPS representation in DMRG \cite{Schollwock2011Density} (for the
ground states), TEBD \cite{Vidal2003Efficient} (for time evolution),
and DMRG-X \cite{Khemani2016Obtaining} (for highly excited eigenstates
of local Hamiltonians deep in the many-body localization region) algorithms.
Yet, the MPS/tensor-network representation
is efficient only in describing quantum states with area-law entanglement
and thus presents serious practical limitations 
in solving problems involving volume-law entanglement states. As introduced
in the previous section, the construction philosophy of neural-network
states is very different from that of MPS/tensor-network states.
This gives rise to the possibility for neural networks to represent
efficiently quantum states and solve problems with volume-law entanglement.
We also stress that our \textit{exact} results here provide an important
anchor point for future theoretical and numerical studies and should
have far-reaching implications in the applications of machine learning
techniques in solving currently intractable many-body problems.  In the subsection C,  we indeed use RBMs to solve the ground state (with massive power-law entanglement) of a modified Haldane-Shastry model with long-range interactions by using the reinforcement learning.

We first give an 1D example. Let us consider an 1D system of $N$ qubits. 
The goal is to  construct a RBM state with maximal volume-law entanglement
entropy. To this end, we introduce a RBM with $N$ visible and $M=\lfloor\frac{3N}{2}\rfloor-1$
hidden neurons. Here, the floor function $\lfloor x\rfloor$ denotes
the largest integer less or equal to $x.$ The weight parameters of
$\Phi_{M}(\Xi;\Omega)$, which characterize the RBM as defined in
Eq. (\ref{eq:RBM states}), are chosen to be 
\begin{eqnarray}
a_{k} & = & 0,\;\forall k\in[1,N],\label{eq:VolLaw1Da}\\
b_{k} & = & \begin{cases}
-\frac{i\pi}{4}, & k\in[1,N-1]\\
\frac{i\pi}{2}, & k\in[N,\lfloor\frac{3N}{2}\rfloor-1]
\end{cases},\label{eq:VolLaw1Db}\\
W_{k'k} & = & \begin{cases}
\frac{i\pi}{4}, & (k',k)\in\mathcal{S}\\
0, & \text{otherwise}
\end{cases},\label{eq:VolLaw1Dc}
\end{eqnarray}
 where $\mathcal{S}$ is a set of paired integers defined by
$\mathcal{S}\equiv\{(i,j):i\in[1,N-1]\;\text{and }j=i,i+1;\;\text{or }i\in[N,\lfloor\frac{3N}{2}\rfloor-1]\;\text{and }j=i+1-N,i+1-\lceil\frac{N}{2}\rceil\}$. 
%{\color{red} [There must be a typo here. ---XL]} 
The ceiling function $\lceil x\rceil$ denotes the smallest integer
greater than or equal to $x.$ A pictorial illustration of this RBM
is shown in Fig. \ref{fig:VolumeLaw1D}. Now, we show that the quantum
states described by the above RBM have volume-law entanglement entropy
for any contiguous region no larger than half of the system size.
To be more precise, we have the following theorem:

\textit{Theorem 2.}\textemdash For an 1D RBM state with weight parameters
specified by Eqs.(\ref{eq:VolLaw1Da}-\ref{eq:VolLaw1Dc}), if we
divide the system into two parts $A$ and $B$ with $A$ consisting
the first $l$ ($1\leq l\leq\lfloor\frac{N}{2}\rfloor$) qubits and
$B$ the rest, then the corresponding R\'{e}nyi entropy of $\rho_{A}$
is
\begin{eqnarray}
S_{\alpha}^{A} & = & l\log2,\quad\forall\alpha.\label{eq:VolLaw}
\end{eqnarray}

\textit{Proof. }As mentioned in Sec. III, since there is no intra-layer
connection between neurons for a RBM, we can explicitly factor out
the hidden variables and rewrite $\Phi_{M}(\Xi;\Omega)$ in a product
form: $\Phi_{M}(\Xi;\Omega)=\prod_{k=1}^{N}e^{a_{k}\sigma_{k}^{z}}\prod_{k'=1}^{M}\Gamma_{k'}(\Xi)$,
with $\Gamma_{k'}(\Xi)=2\cosh(b_{k'}+\sum_{k}W_{k'k}\sigma_{k}^{z})$.
From Eq. (\ref{eq:VolLaw1Da}), $a_{k}=0$ for all $k\in[1,N]$, thus
the first term $\prod_{k=1}^{N}e^{a_{k}\sigma_{k}^{z}}$ simply equals
one and can be omitted from $\Phi_{M}(\Xi;\Omega)$. Consequently,
the variational wavefunction $\Phi_{M}(\Xi;\Omega)$ only depends
on the $\Gamma_{k'}$-factors, which correspond to the hidden neurons.
As shown in Fig. \ref{fig:VolumeLaw1D}, for $k'\in[1,N-1]$, $\Gamma_{k'}=2\cosh[-\frac{i\pi}{4}+\frac{i\pi}{4}(\sigma_{k'}^{z}+\sigma_{k'+1}^{z})]$
connects its corresponding hidden neuron at site $k'$ to two nearest-neighbor
visible neurons at sites $k'$ and $k'+1$. $\Gamma_{k'}$ has only
two possible values: $\Gamma_{k'}=-\sqrt{2}$ if $\sigma_{k'}^{z}=\sigma_{k'+1}^{z}=-1$
and $\Gamma_{k'}=\sqrt{2}$ otherwise. For $k'\in[N,\lfloor\frac{3N}{2}\rfloor-1],$
$\Gamma_{k'}=2\cosh[\frac{i\pi}{2}+\frac{i\pi}{4}(\sigma_{k'+1-N}^{z}+\sigma_{k'+1-\lceil\frac{N}{2}\rceil}^{z})]$
connects its corresponding hidden neuron at site $k'$ to two far-away
separated visible neurons at sites $k'+1-N$ and $k'+1-\lceil\frac{N}{2}\rceil$.
$\Gamma_{k'}$ vanishes if $\sigma_{k'+1-N}^{z}=-\sigma_{k'+1-\lceil\frac{N}{2}\rceil}^{z}$
and $\Gamma_{k'}=2$ ($\Gamma_{k'}=-2$) if $\sigma_{k'+1-N}^{z}=\sigma_{k'+1-\lceil\frac{N}{2}\rceil}^{z}=-1$
($\sigma_{k'+1-N}^{z}=\sigma_{k'+1-\lceil\frac{N}{2}\rceil}^{z}=1$).
These features of the $\Gamma_{k'}$-factors are crucial in the following
proof of Eq. (\ref{eq:VolLaw}). 

For convenience, we define two sets of integers: $\mathcal{B}_1=\{j:N+1-\lceil\frac{N}{2}\rceil\leq j\leq N+l+1-\lceil\frac{N}{2}\rceil\}$
and $\mathcal{B}_2=\{j:l+1<j<N+1-\lceil\frac{N}{2}\rceil\;\text{or }N+l+1-\lceil\frac{N}{2}\rceil<j\leq N\}$.
%{\color{red} [I would like to suggest using ${\cal B}_1$ and ${\cal B}_2$ instead of ${\cal A}$ and ${\cal B}$. ---XL]}
We note that $B=\mathcal{B}_1\cup\mathcal{B}_2$. 
%For a region $X$, we denote the spin configuration within this region as $\Xi_{X}$. 
By using the features of the $\Gamma_{k'}$-factors discussed above,
the RBM state reduces to $|\Psi(\Omega)\rangle\equiv\sum_{\Xi_{A\cup\mathcal{B}_2}}\chi_{\Xi_{A\cup\mathcal{B}_2}}|\Xi_{A}\rangle|\Xi_{\mathcal{B}_1}:\Xi_{\mathcal{B}_1}=\Xi_{A}\rangle|\Xi_{\mathcal{B}_2}\rangle$,
where $\chi_{\Xi_{A\cup\mathcal{B}_2}}=\pm C$ is a coefficient depending
on $\Xi_{A\cup B}$ with $C$ a positive constant.
%{\color{red} [Maybe inserting a delta function say $\delta_{\Xi_{\mathcal{A}} \Xi_{A}}$ is better.---XL ] } 
By tracing out the degrees of freedom in region
$B$ and putting back the normalization constant, we obtain the reduced
density matrix $\rho_{A}=\mathbf{I}/2^{l}$, with $\mathbf{I}$ the
identity matrix of dimension $2^{l}\times2^{l}$. This completes the
proof.

It is worthwhile to mention that the subregion $A$ does \textit{not}
necessarily have to be at the left end. In fact, $A$ can be any contiguous
region of length $l$ and Eq. (\ref{eq:VolLaw}) still remains valid, although
 the details of the proof would change slightly in this situation. We choose $A$ to be at the left
end just for convenience. In the limit $N\rightarrow\infty$, for
any contiguous region its entanglement entropy scales linearly with
the size of the region\textemdash a volume-law entanglement. 

\begin{figure*}
\includegraphics[width=0.98\textwidth]{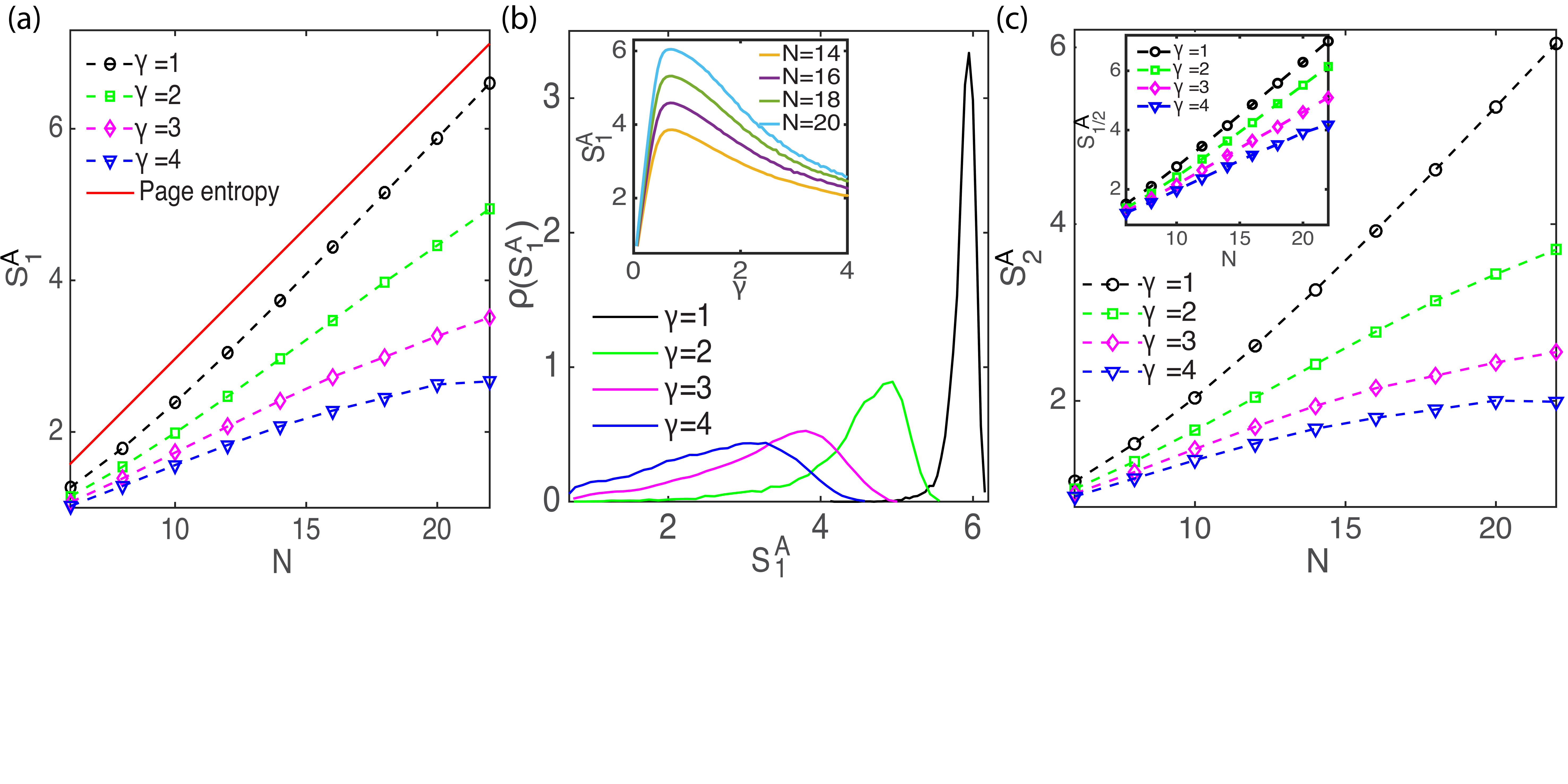}

\caption{Entanglement scaling and distributions of neural-network states with
random weight parameters. (a) The averaged von Neumann entropy $S_{1}^{A}$.
For small $\gamma$ ($\gamma=M/N$ denoting the ratio between the
number of hidden and visible neurons), $S_{1}^{A}$ scales linearly
with the system size, indicating an volume-law entanglement. Yet,
$S_{1}^{A}$ tends to saturate when $\gamma$ becomes large ($\gamma=4$).
Moreover, for all $\gamma$ values studied $S_{1}^{A}$ deviates significantly
from the Page-value for the entanglement entropy averaged over random
pure states. (b) The entanglement distribution of $S_{1}^{A}$
for different $\gamma$. The peak shifts to the left as $\gamma$
increases, which is consistent with the observation that the averaged
$S_{1}^{A}$ decreases as we increase $\gamma$. Here, the system
size is chosen to be $L=20$ and we have used $10^{4}$ random samples.  The inset shows $S_{1}^{A}$ as a function of $\gamma$ for different system sizes. $S_{1}^{A}$ reaches its maximal value at $\gamma^*\approx 0.7$. However, even this maximal value of $S_{1}^{A}$ is still noticeably smaller than the Page value for random states.
(c) The scaling of the R\'{e}nyi entropy with the system size. The second-order
R\'{e}nyi entropy $S_{2}^{A}$ behaves similarly as the von Neumann entropy
$S_{1}^{A}$. The inset shows the results for $S_{1/2}^{A}$. For
simplicity, we have chosen the on-site potentials $a_{k}$ ($k\in[1,N]$)
and $b_{k'}$ ($k'\in[1,M])$ to be zero.  The connection weight parameters
$W_{kk'}$ are complex numbers randomly drawn from uniform distributions with $\text{real}(W_{kk'})\in [-3/N,3/N]$ and $\text{imag}(W_{kk'})\in [-\pi,\pi]$.
\label{fig:Entanglement entropy scaling and distribution}}
\end{figure*}

For the 2D case, we can construct  volume-law entangled RBM states in 
a similar manner. We consider a system of $N$ qubits living on
a $L_{x}\times L_{y}$ square lattice denoted as $\Lambda$. We assume
$L_{x}$ and $L_{y}$ are even integer numbers for simplicity (one
can use the floor and ceiling functions to deal with the case
of odd numbers, but the notations will be more cumbersome). We label
each vertex of the lattice by a pair of indices ($k_{x}$, $k_{y}$)
($1\leq k_{x}\leq L_{x}$ and $1\leq k_{y}\leq L_{y}$) and attach
to it a qubit $N=L_{x}\times L_{y}$. We construct a RBM with $N$
visible and $\frac{5}{2}L_{x}L_{y}-L_{x}-L_{y}$ hidden neurons. The
hidden neurons are divided into three groups. The first (second) group,
denoted by $\mathcal{X}$ ($\mathcal{Y}$), has $(L_{x}-1)\times L_{y}$
($L_{x}\times(L_{y}-1)$) neurons that connect nearest visible neurons
along the $x$ ($y$) direction. The third group, denoted by $\mathcal{Z}$,
contains $L_{x}\times\frac{1}{2}L_{y}$ hidden neurons that connect
visible neurons nonlocally. One can draw an analogy with the 1D example:
the neurons in groups $\mathcal{X}$ and $\mathcal{Y}$ connect nearest
visible neurons and they correspond to the first $N-1$ hidden neurons
in the 1D case, and similarly those in group $\mathcal{Z}$ correspond
to the remaining ones. 
The hidden neurons in ${\cal X}$, ${\cal Y}$ and ${\cal Z}$ are labeled by 
${\bf x} = (x_1, x_2)$, ${\bf y} = (y_1, y_2)$, and ${\bf z} = (z_1, z_2)$, respectively. 
Following the 1D example,
the weight parameters can be chosen as
\begin{eqnarray*}
a_{k_{x}k_{y}} & = & 0,\;\forall(k_{x},k_{y})\in\Lambda,\\
b_{\bf x} ^{({\cal X} )} &=& b_{\bf y} ^{({\cal Y})} = -\frac{i\pi}{4}, \,\,\, 
 b_{\bf z} ^{({\cal Z} )} = \frac{i\pi}{2}  \\  
%b_{k_{x}k_{y}}^{(\mathcal{P})} & = & \begin{cases} -\frac{i\pi}{4}, & \forall(k_{x},k_{y})\in\mathcal{P}\;\text{and }\text{\ensuremath{\mathcal{P}}=\ensuremath{\mathcal{X}},\ensuremath{\mathcal{Y}}}\\
%\frac{i\pi}{2}, & \forall(k_{x},k_{y})\in\mathcal{P}\;\text{and }\mathcal{P}=\mathcal{Z}
%\end{cases},\\
W^{({\cal X}/{\cal Y} /{\cal Z} )} _{{\bf x}/{\bf y}/{\bf z}; k_{x}k_{y}} & = & \begin{cases}
\frac{i\pi}{4}, & ({\bf x}/{\bf y}/{\bf z}; k_{x},k_{y})\in\mathcal{S}_{2D} ^{({\cal X}/{\cal Y} /{\cal Z} )} \\
0, & \text{otherwise}
\end{cases}, 
%W^{X} _{{\bf x}; k_{x}k_{y}} & = & \begin{cases}
%\frac{i\pi}{4}, & (k_{x}',k_{y}',k_{x},k_{y})\in\mathcal{S}_{2D}\\
%0, & \text{otherwise}
%\end{cases},
\end{eqnarray*}
where $\mathcal{S}_{2D} ^{({\cal X} )}$, ${\cal S}_{2D} ^{( {\cal Y})}$, and  ${\cal S}_{2D} ^{ ({\cal Z})}$  are the three  sets that specify the connections between the visible neurons and the three groups of hidden neurons. 
%hidden and visible neurons. 
They are defined as 
$\mathcal{S}_{2D} ^{({\cal X})}\equiv\{({\bf x}; k_{x},k_{y}): k_{x}=x_1,x_1+1; k_{y}=x_2\}$, 
${\cal S}_{2D} ^{({\cal Y})} \equiv \{({\bf y};  k_x, k_y): k_{x}=y_1; k_{y}=y_2,y_2+1\}$, 
$ 
{\cal S}_{2D} ^{({\cal Z})} \equiv \{ ({\bf z}; k_x, k_y): (k_{x},k_{y})=(z_1, z_2),(f(z_1),z_2+\frac{L_{y}}{2})\}$,
with $f(z_1)=z_1 +\frac{L_{x}}{2}$ if $1\leq z_1 \leq\frac{L_{x}}{2}$
and $f(z_1)=z_1 -\frac{L_{x}}{2}$ if $\frac{L_{x}}{2}< z_1\leq L_{x}$.
Following the proof of theorem 2, it is straightforward to verify
that the entanglement entropy for any small regular contiguous subregion
$A$ scales linearly with the volume of $A$ and is maximal $S_{\alpha}^{A}=N_{A}\log2$.
Here, $N_{A}$ denotes the number of qubits inside region $A$. 

We mention that similar constructions carry over to higher dimensions straightforwardly. For a system defined on a simple cubic lattice in $d$-dimension with $N=L^d$ qubits, our construction requires  $M=\frac{2d+1}{2}L^d-dL^{d-1}$ hidden neurons and $3M$ nonzero weight parameters. Both the number of hidden neurons and the number of parameters scale only linearly   with the system size. In contrast, if we express these RBM states in terms of  MPS/tensor-networks,  the bond dimension will grow exponentially with the system size, and the problem quickly becomes intractable. This demonstrates explicitly a unique advantage of RBMs in representing quantum many-body states with massive entanglement.

\subsection{Entanglement benchmarking }

%In the above section, we proved that all short-range RBM states satisfy an area law of entanglement. What about RBM states with long-range connections? 
For a general RBM state with long-range connections, the entanglement entropy cannot be calculated analytically. 
%analytically calculating its entanglement entropy is extremely tough or even not possible. 
We thus resort to numerical simulations. We
study the entanglement properties of RBM states with random weight
parameters.  We consider an 1D system with $N$ qubits. The corresponding
RBM has $N$ visible and $M$ hidden neurons with the weight parameters
chosen randomly and independently.  For each random sample, we calculate numerically the coefficients for all possible spin configurations (there are $2^N$ configurations) and normalize them to obtain the corresponding quantum state in the computational basis. We then make an equal bipartition and calculate the reduced density matrix $\rho_A$ for the A subsystem.  We diagonalize $\rho_A$ to compute the desired entanglement entropy and spectrum.
%For each random sample, we make
%an equal bipartition and compute the corresponding entanglement entropy
%and spectrum by exact numerical method. 
The number of samples used
for numerics ranges from $10^{6}$ $(N=6)$ to $10^{3}$ ($N=22$). We mention
that although we focus only on 1D systems, some entanglement features
discovered here should carry over to higher dimensions as well. Extensive higher dimensional RMB-based numerics are left for future studies.

In Fig. \ref{fig:Entanglement entropy scaling and distribution}(a),
we plot the averaged entanglement entropy scaling with different system sizes.
When $\gamma$ is small ($\gamma=1,2,3$), we find that the averaged
entanglement entropy scales linearly with the system size\textemdash a
volume law  (This is another indication that entanglement is not the limiting factor for the RBMs in representing quantum many-body states). Here, $\gamma=M/N$ denotes the ratio between the number
of hidden and visible neurons.  However, when $\gamma$ increases the
entanglement apparently bends downwards and seems to saturate at large
$N$. This appears surprising at the first sight because an increase
of $\gamma$ means an increasing of number of connections between
visible neurons, and intuitively the entanglement should increase as
well. In fact, the bending of the curve at large $\gamma$ may be understood
by looking at the original RBM representation in Eq. (\ref{eq:RBM states}).
Since we choose $W_{kk'}$ randomly, on average $\Phi_{M}(\Xi;\Omega)$
will become less and less dependent on the spin configuration $\Xi$
as $\gamma$ increases. In other words, in the represented many-body
quantum wavefunction the difference between the coefficients of each
component becomes smaller and smaller. Thus, the state become closer
and closer to a product state and therefore the entanglement decreases.  This is further justified in the inset of Fig. \ref{fig:Entanglement entropy scaling and distribution}(b), where the von Neumann entropy $S^A_1$ is shown as a function of $\gamma$ for different system sizes. From this figure, $S^A_1$ reaches its maximal value at a critical  $\gamma^*\approx 0.7$, independent of system size. When $\gamma>\gamma^*$, $S^A_1$ decreases as we increase $\gamma$. It is also worthwhile to mention that when we fix $M$ as a finite number (then $\gamma\rightarrow 0$ in the thermodynamic limit $N\rightarrow \infty$), then $S^A_1$ is upper bounded by $M\log2$, regardless of the system size. This can be understood heuristically by imagining all the hidden neurons being grouped into the  subsystem B, then the subsystem A can only have at most $2^M$ degrees of freedom that are entangled with B. Consequently, $S^A_1$ is bounded by $M\log2$. This explains the numerical observation in the inset of Fig. \ref{fig:Entanglement entropy scaling and distribution}(b) that for $M=1$ (smallest $\gamma$), $S^A_1\approx \log2$ independent of the system size.

\begin{figure*}
\includegraphics[width=0.98\textwidth]{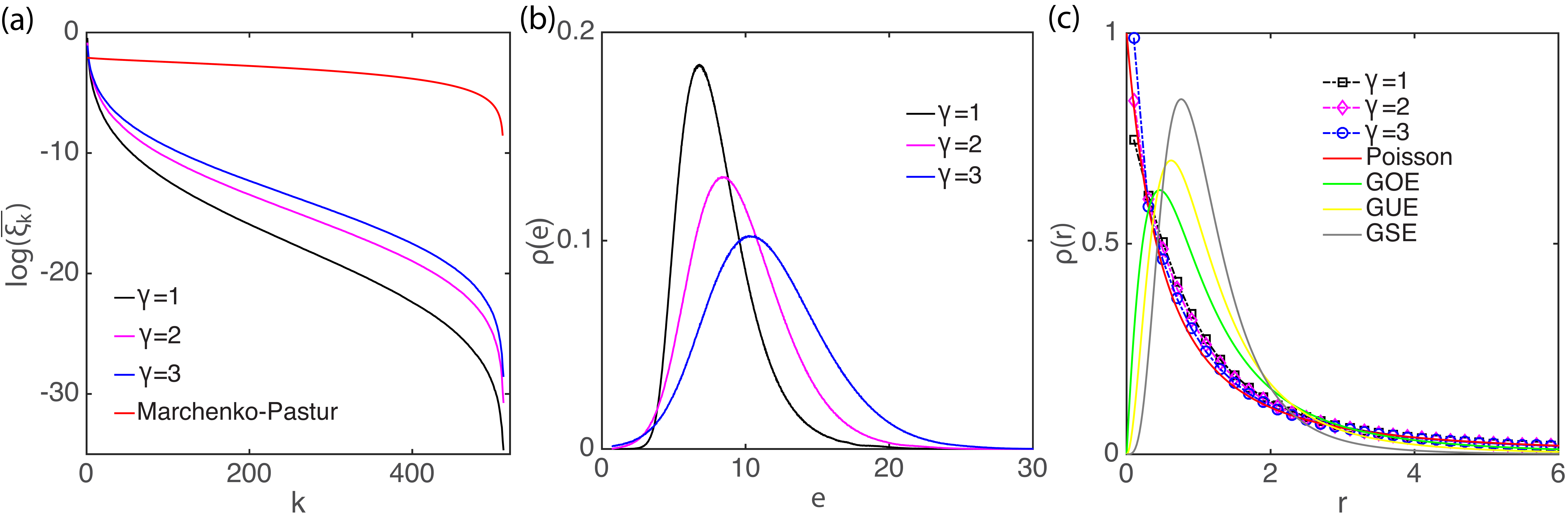}

\caption{Entanglement spectrum and level statistics for neural-network states
with random weight parameters. Here, the lattice size is chosen to
be $L=20$ and we have used $10^{4}$ random samples.  The random parameters are drawn from the same distribution as specified in Fig. \ref{fig:Entanglement entropy scaling and distribution}. (a) Averaged
entanglement spectrum with different $\gamma$. $\xi_{k}$ denotes
the $k$-th eigenvalue of $\rho_{A}$ (the eigenvalue is arranged
in descending order). The red line denotes the spectrum of a completely
random state (derived from a Marchenko-Pastur distribution). It is
evident that the entanglement spectra of the restricted-Boltzmann-machine
states with random parameters are completely distinct from the Marchenko-Pastur
distribution, indicating that their corresponding entanglement Hamiltonian
are very different from Wishart matrices. (b) Density of states for
the entanglement Hamiltonians. Here, $e$ denotes the eigenspectrum
of $H_{\text{ent}}$. When increasing $\gamma$, the distribution
of the eigenvalues broadens and the peak shifts rightwards. (c) Distributions
of the ratios of consecutive spacings for the entanglement spectrum.
These distributions follow a Poisson law and differs significantly
from the predictions of GOE, GUE or GSE. \label{fig:Entanglement-spectrum}}
\end{figure*}

In order to compare the RBM states with random parameters with generic
random pure states, we also calculate the so-called Page entropy \cite{Page1993Average},
which is the averaged entanglement entropy over pure states drawn
randomly from the entire Hilbert space of the system. The Page entropy
provides an estimate for entanglement in extended thermal states
\cite{Khemani2016Critical} and has been widely used in the context
of quantum chaos \cite{Monasterio2005Entanglement}, blackhole information
\cite{Harlow2016Jerusalem,Page1993Information}, and many-body localization
\cite{Kjall2014Many,Li2016Statistical}. From the random matrix theory,
it can be computed as: $S_{\text{Page}}=-\frac{d_{A}-1}{2d_{B}}+\sum_{j=d_{B}+1}^{d_{A}d_{B}}\frac{1}{k}$,
where $d_{A}$ and $d_{B}$ denote the Hilbert space dimensions of subsystems
$A$ and \textbf{$B$}, respectively \cite{Page1993Average}.  An interesting
observation in Fig. \ref{fig:Entanglement entropy scaling and distribution}(a) and the inset of Fig. \ref{fig:Entanglement entropy scaling and distribution}(b)
is that the entanglement entropy is always smaller than the Page entropy
for all $\gamma$. This implies that the pattern of entanglement for
the RBM states with random parameters is distinct from that of random
pure states, which is consistent with the fact that the RBM states
live in a very small restricted subspace of the entire Hilbert space. This also indicates that a random state in the Hilbert space is probably not well described by a RBM efficiently.
In Fig. \ref{fig:Entanglement entropy scaling and distribution}(b),
we plot the entanglement distribution for different $\gamma$. We
find that, as $\gamma$ increases, the distribution becomes broader
and the density peak shifts towards smaller values. This is in agreement
with the observation in Fig. \ref{fig:Entanglement entropy scaling and distribution}(a)
that the entanglement decreases as $\gamma$ increases. Fig. \ref{fig:Entanglement entropy scaling and distribution}(c)
shows the results for the R\'{e}nyi entropy of orders two and one half.
As expected, $S_{2}$ behaves very similarly to $S_{1}$. For $S_{1/2}$,
we find a similar volume-law scaling of entanglement, but the bending 
feature does not show up at $\gamma=4$ due to finite-size effects.

The entanglement entropy studied above provides a wealth of information
about the data structure of the RBM states. Yet, as has been realized
in a number of different physical contexts, the entanglement entropy
cannot capture the full entanglement structure of the system \cite{Susskind2016Entanglement,Yang2005Two,Geraedts2016Many,Cirac2011Entanglement,Calabrese2008Entanglement,Garrison2015Does}.
Much greater information can be extracted from the entanglement spectrum.
In order to obtain a more comprehensive understanding of the data
structure of the RBM states with random weight parameters, we have therefore also
calculated their entanglement spectra and the entanglement
Hamiltonian level statistics. In Fig. \ref{fig:Entanglement-spectrum}(a),
we plot the averaged entanglement spectrum for different $\gamma$.
We find that the entanglement spectrum for the RBM states is completely
different from the Marchenko-Pastur distribution derived from random
matrix theory \cite{Marvcenko1967Distribution}. More specifically,
the Marchenko-Pastur distribution describes the asymptotic average
density of eigenvalues of a Wishart matrix (a matrix of the form $Y=XX^{\dagger}$
with $X$ a random rectangular matrix). It is shown recently in Ref.
\cite{Yang2005Two} that the entanglement spectrum of highly excited
eigenstates in the delocalized phase bears a two-component structure:
(i) a universal part that is associated with random matrix theory,
i.e., a universal tail that follows the Marchenko-Pastur distribution
and thus is model independent, and (ii) a model-dependent nonuniversal
part which dominates the weights in the spectrum. In the localized
phase, the universal part of the spectrum disappears in the thermodynamic
limit, leaving only the nonuniversal part that leads to an area-law
scaling of the entanglement entropy. In our case for the RBM states
with random weight parameters, the universal part disappears completely
even for a system size as small as $N=20$. In this sense, these RBM
states are less random than the highly excited eigenstates in both
the delocalized and localized phases. This further shows that, although
these RBM states obey a volume law  of entanglement entropy on average,
they are living in a very restricted subspace of the entire Hilbert
space. In Fig. \ref{fig:Entanglement-spectrum}(b), we plot the density
of states for the entanglement Hamiltonian of these RBM states. We
find that a broader distribution shows up as we increase $\gamma$
and the peak moves to the right, which is consistent with the surprising
results from Fig. \ref{fig:Entanglement entropy scaling and distribution}
(i.e., entanglement decreases as $\gamma$ increases).

Another quantity which is also useful in understanding the data structure
of the RBM states is the adjacent gap ratio $r$ defined as $r_{n}=\frac{\min(\delta_{n},\delta_{n-1})}{\max(\delta_{n},\delta_{n-1})}$,
with $\delta_{n}$ the level spacing between the $n$-th and the $(n-1)$-th
eigenstates of the entanglement Hamiltonian. We note that the importance
of the distribution of $r$ has been broadly appreciated in various
contexts. In quantum chaos \cite{Brody1981Random}, it is argued that
whereas the level statistics for integrable quantum Hamiltonians obeys
a Poisson law \cite{Berry1977Level}, the case for Hamiltonians with
chaotic dynamics must follow one of the three classical ensembles
from random matrix theory \cite{Bohigas1984Characterization}, namely
the Gaussian orthogonal ensemble (GOE), the Gaussian unitary ensemble
(GUE) and the Gaussian symplectic ensemble (GSE). These three ensembles
correspond to Hermitian random matrices with entries being independently
distributed random real, complex, and quaternionic variables, respectively
\cite{Akemann2011Oxford}. In many-body localization, it is generally
believed (verified by extensive numerical calculations recently) that
Hamiltonians in the delocalized and localized regions manifest respectively
GOE (or GUE) and Poisson level statistics \cite{Pal2010Manybody}.
The level statistics of the entanglement Hamiltonians in this context
has also been studied recently in Ref. \cite{Geraedts2016Many}. It
was shown that in the thermal phase the entanglement spectrum
shows level statistics that in agreement with predictions from random
matrix theory and is governed by the same random matrix ensemble as
the energy spectrum. Yet, in the many-body localized phase, the entanglement
spectrum shows a semi-Poisson distribution, in contrast to the energy
spectrum following a Poisson law. For the RBM states with random
weight parameters studied in this section, we find that their entanglement
spectra follow a Poisson distribution, as shown in Fig. \ref{fig:Entanglement-spectrum}(c).
The averaged value of $r_{n}$ (over $10^{4}$ random samples) with
$N=20$ and $\gamma=3$ equals $0.378$, which is in good agreement
with the Poisson predicted value $2\ln2-1\approx0.386$ \cite{Atas2013Distribution}.
The small deviation could be attributed to finite-size effect. Thus,
these RBM states are distinct from the eigenstates of Hamiltonians
in either the delocalized or localized phases on average. In addition,
we also remark that the Poisson behavior and the lack of universal
part in the entanglement spectra of these RBM states imply that
they are \textit{not} irreversible\textemdash namely there exists
an efficient algorithm to completely disentangle these states \cite{Chamon2014Emergent}.
Finding out the disentangling algorithm would provide some  insight
into the nature of neural-network quantum states and is an interesting
topic for future investigation.

\subsection{Reinforcement learning of ground states with power-law entanglement}
\begin{figure}
\includegraphics[height=0.33\textwidth]{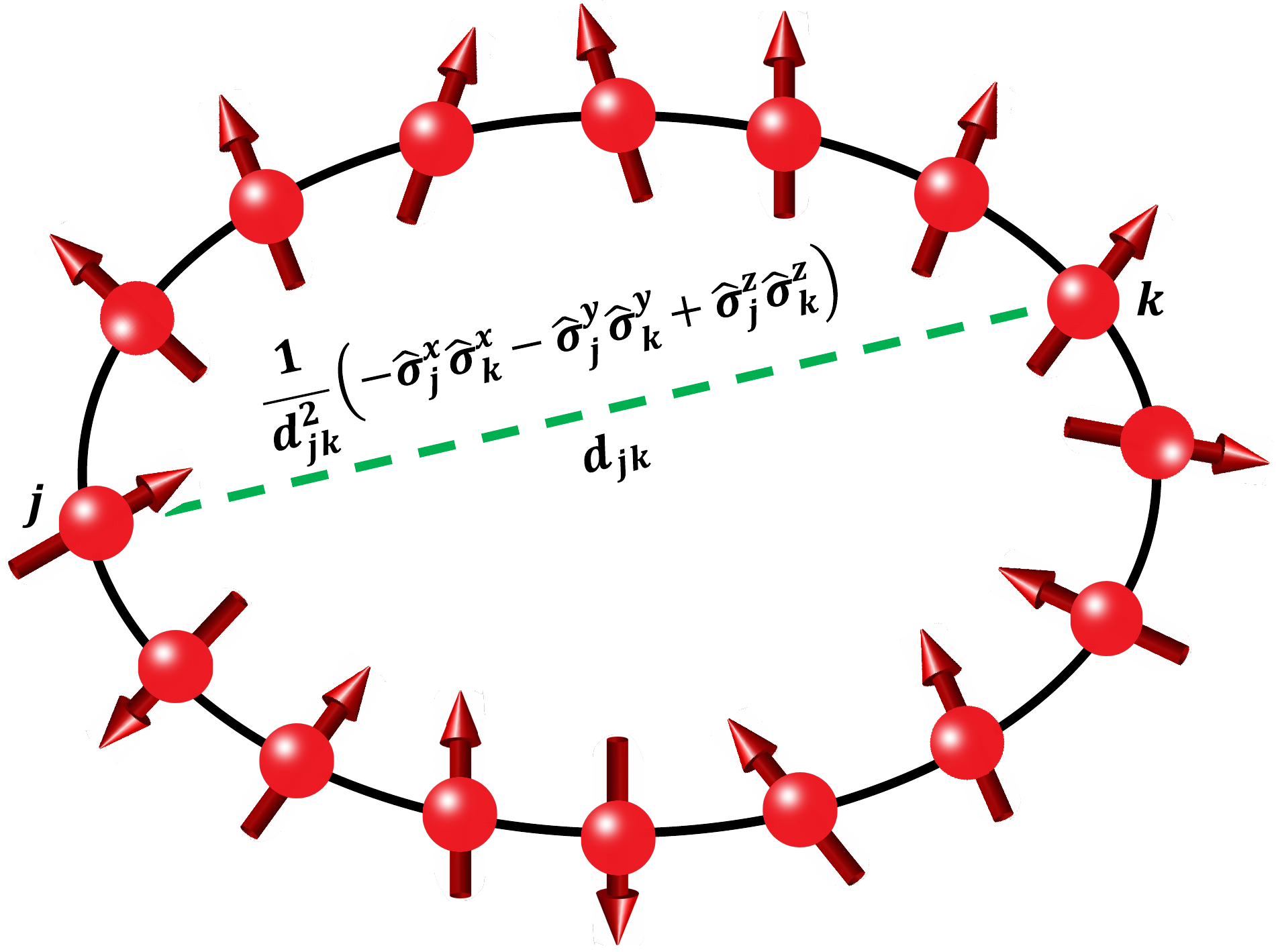}
\caption{ The modified Haldane-Shastry model. $N$ spin-$\frac{1}{2}$ particles form an equally-spaced lattice on a ring. Each spin is interacting with all the other spins. The interaction between spin $j$ and $k$ is of Heisenberg XXZ type and its strength is inversely proportional to the square of the chord distance $d_{jk}$ (see Eq. (\ref{Eq: HamHS}) in the main text). The ground state of this model has power-law entanglement and can be calculated with restricted Boltzmann machine through reinforcement learning.
\label{Fig:HSMModel}}
\end{figure}

\begin{figure}
\includegraphics[height=0.491\textwidth]{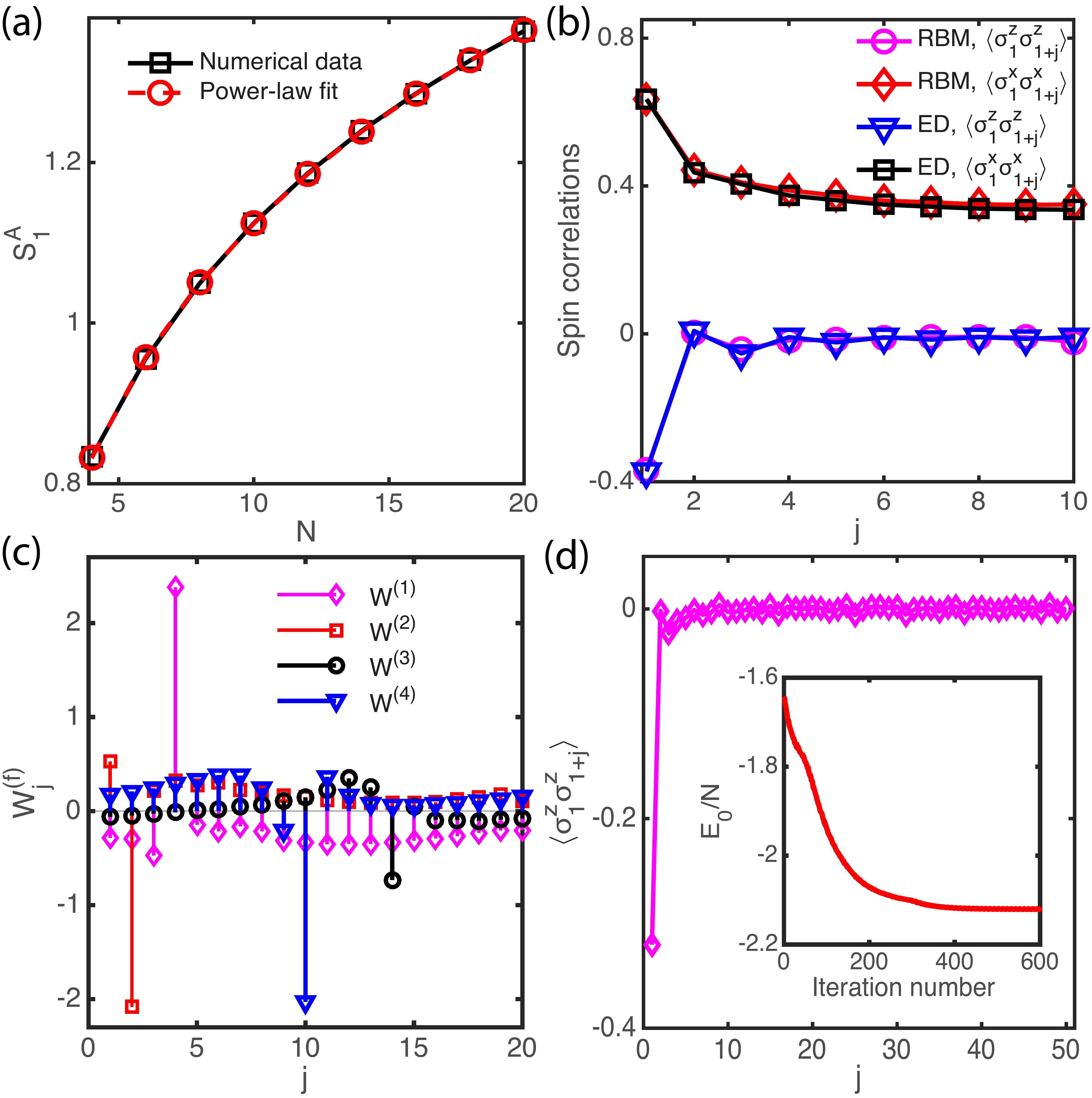}
\caption{ Reinforcement learning of the ground state of the modified Haldane-Shastry model with power-law entanglement. (a) The von Neumann entropy calculated by exact diagonalization (ED) of the Hamiltonian with different system sizes.  It has an excellent power-law fit: $S^A_1\sim 2.473N^{0.106}-2.033$. Here, the entropy is calculated from an equal bipartition of system. (b) Spin correlations for small system sizes calculated by ED and restricted Boltzmann machine, respectively. (c) The learned feature maps for representing the ground state with restricted Boltzmann machine. In (b) and (c), the hidden-unit density $\gamma=4$ and the system size is fixed to $N=20$. (d) Reinforcement learning of the spin correlations and ground state energy density for larger system size $N=100$. The inset shows the variational  energy density as a function of the iteration number of the learning process. As the machine learns more and more round, the energy converges smoothly to an  asymptotic value around $E_0/N\approx -2.12$. 
\label{Fig:HSMEDRBM}}
\end{figure}

The above discussion shows that, unlike MPS/tensor-network states, entanglement is not the limiting factor for the efficiency of the RBM representation. As an important consequence, RBM might be capable of solving some quantum many-body problems where massive entanglement is involved. To demonstrate this unprecedented power, in this section we consider the problem of finding the ground state (with power-law entanglement) of a spin-$1/2$ Hamiltonian with long-range interaction, through a reinforcement-learning scheme \cite{Sorella2007Weak,Carleo2016Solving}. We consider $N$ spin-$1/2$ particles living on a ring (see Fig. \ref{Fig:HSMModel}) with a modified Haldane-Shastry  \cite{Haldane1988Exact,Shastry1988Exact} Hamiltonian given by 
\begin{eqnarray}
H_{\text{MHS}}=\sum_{j<k}^N \frac{1}{d_{jk}^2}(-\hat{\sigma}_i^x\hat{\sigma}_j^x-\hat{\sigma}_i^y\hat{\sigma}_j^y+\hat{\sigma}_i^z\hat{\sigma}_j^z), \label{Eq: HamHS}
\end{eqnarray}
where $d_{ij}=\frac{N}{\pi}|\sin[\pi(j-k)/N]|$ is the so-called ``chord distance". %This model differs from the conventional Haldane-Shastry \cite{Haldane1988Exact,Shastry1988Exact} by a minus sign in front of $\hat{\sigma}_i^x\hat{\sigma}_j^x$ and $\hat{\sigma}_i^y\hat{\sigma}_j^y$.  
%{\bf (It may not be a good idea to emphasize too much about the difference. Just saying a modified Haldane-Shastry model should be fine)} 
Since it has long-range interactions with a power-law decaying strength, we expect its ground state to have power-law entanglement.  Although a rigorous proof is still lacking and seems very hard to obtain, we can verify the entanglement power law numerically. In Fig. \ref{Fig:HSMEDRBM}(a), we plot the von Neumann entropy for the ground state of $H_{\text{MHS}}$ calculated from exact diagonalization (ED). We find that it indeed has an excellent power-law fit with the system size.

We now show that RBM is capable of faithfully and efficiently representing the ground state of $H_{\text{MHS}}$ and the  representing RBM can be efficiently obtained via reinforcement learning,  despite the fact that the ground state has a large amount of  entanglement entropy. Since the Hamiltonian has a lattice translation symmetry, we can use this symmetry to reduce the number of variational parameters and for integer hidden-variable density ($\gamma=1,2,\cdots$) the weight matrix takes the form of feature filters $W_j^{(f)}$ with $f\in [1,\gamma]$, as described in Ref. \cite{Carleo2016Solving}. %By using the reinforcement-learning scheme proposed in Ref. \cite{Carleo2016Solving}, we optimize the parameters of the RBM to best represent 
In Fig. \ref{Fig:HSMEDRBM}(b), we plot the different spin correlations obtained via reinforcement learning, for small system sizes. We compare the RBM result with that from exact diagonalization. As shown in this figure, the RBM result matches the ED result very well. The accuracy of the RBM result  can be improved  by increasing $\gamma$ and the number of iterations in the training process. In Fig. \ref{Fig:HSMEDRBM}(c), we show the feature maps after a typical reinforcement learning process with $\gamma=4$ and $N=20$. The accuracy of the trained RBM can be quantified by the relative error on the ground state energy $\epsilon_{\text{rel}}=|(E_0^{(\text{RBM})}-E_0^{\text{ED}})/E_0^{\text{ED}}|$ \cite{Carleo2016Solving}. For the parameters shown in Fig. \ref{Fig:HSMEDRBM}(c), we find $\epsilon_{\text{rel}}\sim 10^{-5}$. We then move on to calculate the correlation functions and ground state energy density for larger system sizes, which are far beyond the capability of the ED technique. We plot some of the results for $N=100$ in Fig. \ref{Fig:HSMEDRBM}(d). We find that the correlation $\langle\hat{\sigma}^z_1\hat{\sigma}^z_{1+j}\rangle$ has a sharp jump at $j=2$, which is also obtained in our ED calculations for smaller system sizes, as shown in Fig.\ref{Fig:HSMEDRBM}(b).

 %different from the result for the conventional Haldane-Shastry model where the corresponding correlation has a power-law decay for large $N$ limit \cite{Gebhard1987Correlation}: $\langle\hat{\sigma}^z_1\hat{\sigma}^z_{1+j}\rangle=\frac{-1}{\pi j} \int_0^j \frac{\sin(\pi x)}{\pi x)} dx$. We mention that this jump feature of $\langle\hat{\sigma}^z_1\hat{\sigma}^z_{1+j}\rangle$ for the modified Haldane-Shastry model is also obtained in our ED calculations for smaller system sizes, as shown in Fig.\ref{Fig:HSMEDRBM}(b).  

%We remark that the DMRG/MPS method is not suitable for the above problem due to 
%the massive power-law entanglement of the ground state, all-to-all long-range interaction, and the ring structure of  $H_{\text{MHS}}$. In this regard, the reinforcement-learning based RBM technique has apparent advantages when large entanglement and long-range interactions are involved. Moreover, as pointed out in Ref. \cite{Carleo2016Solving}, the RBM approach works as well in higher dimensions and for dynamical problems. We also mention that $H_{\text{MHS}}$ might be realized with trapped ions in a ring geometry \cite{Grass2014Trapped,Gong2013Quantum}. Thus the numerically calculated correlations could be experimentally verified in the future. 

 We remark that the DMRG/MPS-based simulations are particularly challenging for the above problem and would presumably require a substantially larger number of variational parameters than the RBM approach \cite{Schollwock2011Density,Haegeman2016Unifying}. In this regard, the reinforcement-learning based RBM technique has apparent advantages when large entanglement and long-range interactions are involved. Moreover, as pointed out in Ref. \cite{Carleo2016Solving}, the RBM approach works as well in higher dimensions and for dynamical problems. We also mention that $H_{\text{MHS}}$ might be realized with trapped ions in a ring geometry \cite{Grass2014Trapped,Gong2013Quantum}. Thus the numerically calculated correlations could be experimentally verified in the future.

%\section{a RBM method to analytically compute entanglement entropy and spectrum}

\section{an analytical RBM recipe for calculating entanglement}
\label{sec:EntSpectrum}
%{\color{red} [There are several places you noted both ``entropy" and ``spectrum". I would suggest just keeping spectrum, since entropy is related to spectrum in a very straightforward way. ---XL]}

In Sec.~\ref{sec:AreaLaw}, we proved that all short-range RBM states obey an area
law entanglement. Can we calculate the entanglement entropy and spectrum
analytically? For a general many-body state, this is an outstanding
challenge, especially for a system with a finite size. In fact, most
of the past works focus on the thermodynamic limit and compute
entanglement entropy asymptotically. The methods used often involve
complicated mathematics \cite{Eisert2010Area}. For instance, the
Fisher-Hartwig formula has been used to evaluate the asymptotic behavior
of the entanglement entropy for the critical XX model and other isotropic
models \cite{Jin2004Quantum,Its2005Entanglement,Eisert2005Single,Keating2005Entanglement}.
Another notable approach is the use of conformal field theory, where
some universal properties of entanglement entropy have been established
for critical $(1+1)$-dimensional systems \cite{Holzhey1994Geometric,Calabrese2004Entanglement,Callan1994Geometric}.
For calculating the exactly entanglement entropy of a finite system, we
note the quotient group method, which has been used to calculate the
entropy of an arbitrary bipartition of the 2D toric code states \cite{Hamma2005Bipartite,Hamma2005Ground}.
Here, however, we introduce an alternative approach and show that the RBM representation
would also help the analytical calculation of the entanglement entropy and
spectrum. As an example, we consider the 1D SPT cluster state $|\Psi\rangle_{\text{cluster}}$,
which is the ground state of the cluster Hamiltonian $H_{\text{cluster}}$
defined on an 1D lattice with periodic boundary condition: $H_{\text{cluster}}=-\sum_{k=1}^{N}\hat{\sigma}_{k-1}^{z}\hat{\sigma}_{k}^{x}\hat{\sigma}_{k+1}^{z}$.
This state is a topological state protected by $Z_{2}\times Z_{2}$
symmetry \cite{Son2012Topological}. It serves as a simple toy model
for studying SPT phases and has important applications in measurement-based
quantum computation \cite{Hein2004Multparty,Raussendorf2003Measurement,Nielsen2006Cluster}.
An exact and efficient RBM representation of the 1D cluster state
has been found in \cite{Deng2016Exact}. This representation has $N$
hidden neurons with each one connecting only locally to the visible
neurons within distance one. The weight parameters are specified as the
following:
\begin{eqnarray}
a_{k} & = & 0,\;b_{k}=\frac{i\pi}{4},\;W_{kj}=\begin{cases}
i\omega_{\mu}, & \text{if }|k-j|=1\\
0, & \text{otherwise}
\end{cases},\label{eq:RBMclusterPara}
\end{eqnarray}
where $\omega_{\mu}$s ($\mu=1,0,-1$) are positive real numbers giving
by $(\omega_{1},\omega_{0},\omega_{-1})=\frac{\pi}{4}(2,3,1)$. In
the product form, the normalized 1D cluster state reads
\begin{eqnarray}
|\Psi\rangle_{\text{cluster}} & = & \sum_{\Xi}\prod_{k'=1}^{N}\Gamma_{k'}(\Xi)|\Xi\rangle,\label{eq:1DclusterSt}
\end{eqnarray}
where the $\Gamma_{k'}$-factor only depends on the configurations
of three nearest visible neurons $\Gamma_{k'}(\Xi)=\Gamma_{k'}(\sigma_{k'-1}^{z},\sigma_{k'}^{z},\sigma_{k'+1}^{z})=\cos[\frac{\pi}{4}(1+2\sigma_{k'-1}^{z}+3\sigma_{k'}^{z}+\sigma_{k'+1}^{z})]$
(note that we have put back the normalization constant and have rescaled all the $\Gamma_{k'}$-factor by $1/2$). 

\begin{figure}
\includegraphics[height=0.41\textwidth]{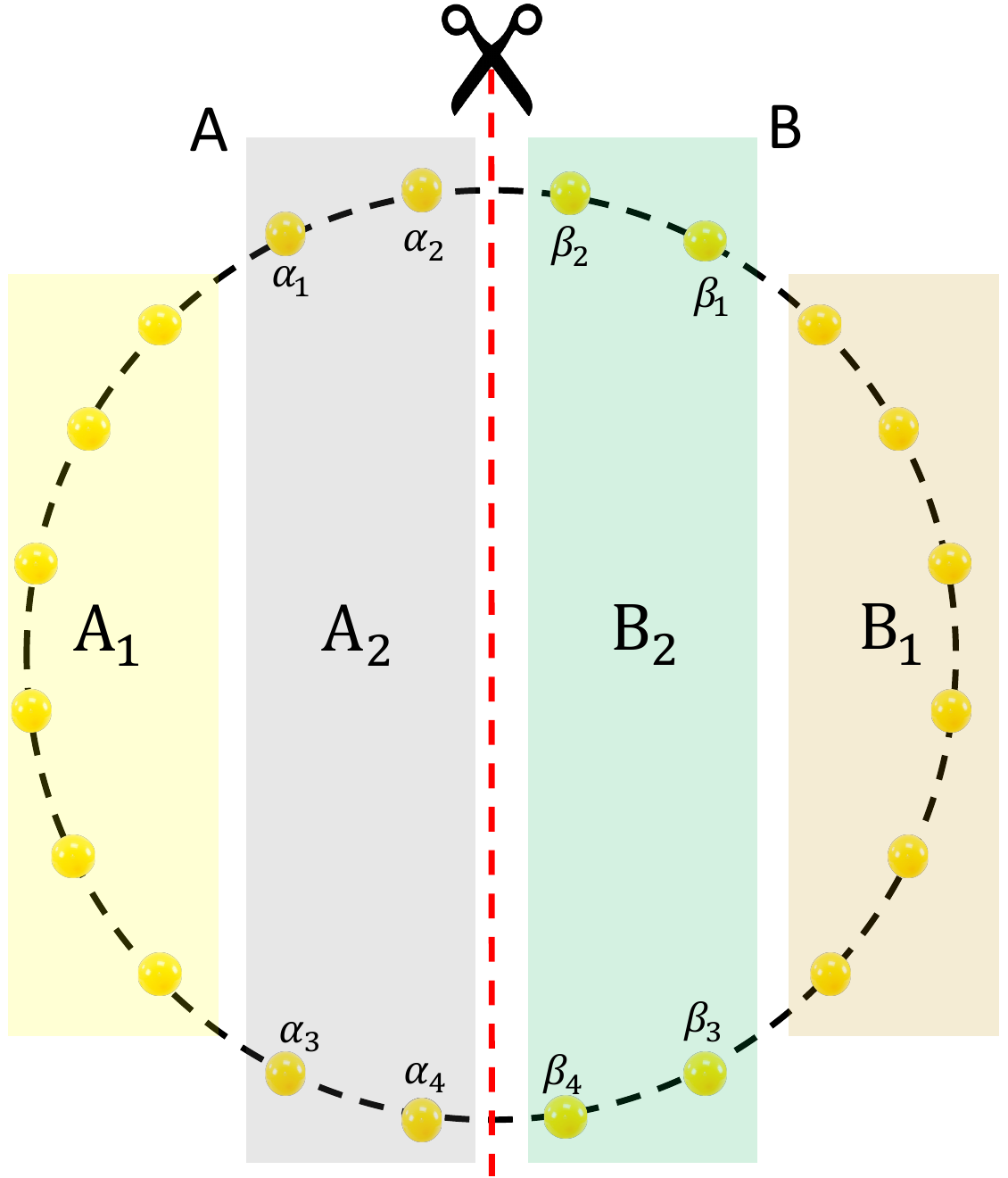}

\caption{A sketch for computing analytically the entanglement entropy and spectrum
of the 1D symmetry-protected-topological cluster state, through the
corresponding exact short-range restricted-Boltzmann-machine representation.
The scissors show a cut of the system into two subsystems $A$ and
$B$. We calculate the entanglement entropy and spectrum from the reduced
density matrix $\rho_{A}$. In order to conveniently explore the short-range
feature of the restricted-Boltzmann-machine, we further divide $A$
($B$) into $A_{1}$ and $A_{2}$ ($B_{1}$ and $B_{2}$). It is important
that the $\Gamma_{k'}$-factors in subregion $A_{1}$ ($B_{1}$) are
independent of spin configurations in $B$ ($A$), such that the summations
of spin configurations in $A_{1}$ and $B_{1}$ are interchangable
and can be factorize out explicitly, as shown in Eq. (\ref{eq:1dstAB}).
\label{Fig:OneDclusterDiv}}
\end{figure}

In order to study its entanglement properties, we consider an arbitrary
bipartition of the system into two parts $A$ and $B$ and we aim
to calculate the entanglement entropy and spectrum of subsystem $A$
analytically. For convenience, we further divide the subregion $A$
($B$) into two parts $A_{1}$ and $A_{2}$ ($B_{1}$ and $B_{2})$
with $A_{2}$ ($B_{2}$) containing only four sites $\alpha_{1}$,
$\alpha_{2}$, $\alpha_{3}$, and $\alpha_{4}$ ($\beta_{1}$, $\beta_{2}$,
$\beta_{3}$, and $\beta_{4}$), as shown in Fig. \ref{Fig:OneDclusterDiv}.
Using the fact that the RBM is short-ranged and $\Gamma_{k'}(\Xi)=\pm\frac{\sqrt{2}}{2}$,
we can rewrite $|\Psi\rangle_{\text{cluster}}$ in the following form,
where the subregions $A$ and $B$ show up explicitly
\begin{eqnarray}
|\Psi\rangle_{\text{cluster}} & =\frac{1}{4} & \sum_{\Xi_{A_{2}\cup B_{2}}}\Gamma_{\alpha_{2}}\Gamma_{\alpha_{4}}\Gamma_{\beta_{2}}\Gamma_{\beta_{4}}|\Psi_{A}\rangle|\Psi_{B}\rangle,\label{eq:1dstAB}
\end{eqnarray}
with $|\Psi_{A}\rangle=|\Psi_{A}(\Xi_{A_{2}})\rangle=2\sum_{\Xi_{A_{1}}}\Gamma_{\alpha_{1}}\Gamma_{\alpha_{3}}\prod_{k'\in A_{1}}\Gamma_{k'}(\Xi_{A})|\Xi_{A}\rangle$
and $|\Psi_{B}\rangle=|\Psi_{B}(\Xi_{B_{2}})\rangle=2\sum_{\Xi_{B_{1}}}\Gamma_{\beta_{1}}\Gamma_{\beta_{3}}\prod_{k'\in B_{1}}\Gamma_{k'}(\Xi_{B})|\Xi_{B}\rangle$.
Eq. (\ref{eq:1dstAB}) is crucial in calculating the entanglement
entropy and spectrum. Compared with Eq. (\ref{eq:1DclusterSt}), it
contains only $2^{8}$, rather than $2^{N}$, terms in the summation.
Noting that $|\Psi_{A}\rangle$ ($|\Psi_{B}\rangle$) only depends
on the spin configurations within subregion $A_{2}$ ($B_{2}$) and
$\langle\Psi_{A}(\Xi_{A_{2}})|\Psi_{A}(\Xi_{A_{2}}')\rangle=\delta_{\Xi_{A_{2}},\Xi_{A_{2}}'}$,
one can do an unitary transformation $U_{A}$ $(U_{B}$) within subregion
$A$ ($B$) to rotate the basis of the Hilbert space $H_{A}$ ($H_{B}$)
of $A$ ($B$). Note that this rotation will not affect the entanglement
entropy and spectrum. In the new basis, $|\Psi_{A}(\Xi_{A_{2}})\rangle$
($|\Psi_{B}(\Xi_{B_{2}})\rangle$) is just a basis vector of $H_{A}$
($H_{B}$). By tracing out the degrees of freedom in subregion $B$
and plugging in the parameter values in Eq. (\ref{eq:RBMclusterPara}),
we find a very simple expression for the reduced density matrix $\rho_{A}$
in the new basis
\begin{eqnarray}
\rho_{A} & = & M_{1}\oplus M_{1}\oplus M_{2}\oplus M_{2}\oplus\mathbf{0},\label{eq:RhoA}
\end{eqnarray}
where $M_{1}=\frac{1}{4}|\psi_{1}\rangle\langle\psi_{1}|$ and $M_{2}=\frac{1}{4}|\psi_{2}\rangle\langle\psi_{2}|$
are four-by-four matrices with $|\psi_{1}\rangle=\frac{1}{2}(|0\rangle+|1\rangle)\otimes(|0\rangle-|1\rangle)$
and $|\psi_{2}\rangle=\frac{1}{2}(|0\rangle-|1\rangle)\otimes(|0\rangle-|1\rangle)$,
respectively; $\mathbf{0}$ is a zero matrix of dimension $2^{N_{A_{1}}}\times2^{N_{A_{1}}}$
with $N_{A_{1}}$ denoting the number of spins in subregion $A_{1}$.
From Eq. (\ref{eq:RhoA}), the eigenvalues of $\rho_{A}$ can be obtained
readily. $\rho_{A}$ has only four nonzero eigenvalues that are degenerate
and equal to $\frac{1}{4}$. As a result, the R\'{e}nyi entropy is given
by
\begin{eqnarray*}
S_{\alpha}^{A} & = & 2\log2,\quad\forall\alpha.
\end{eqnarray*}
For the entanglement spectrum, $H_{\text{ent}}$ has a four-fold degeneracy
with the four smallest eigenvalues equal to $2\log2$ and the rest are
infinite. The four-fold degeneracy is a signature of SPT phases \cite{Li2008Entanglement,Chardran2011Bulkedge,Fidkowski2010Entanglement,Qi2012General}. 

We expect that this RBM approach would carry over to calculating the
entanglement entropy and spectrum for the 2D   and 3D toric code states, whose
RBM representation has already been given in Ref. \cite{Deng2016Exact},
although the calculation will be more technically involved. Undoubtedly,
like all analytical methods in calculating entanglement,  our RBM approach
has obvious limitations and cannot be applied in general to an arbitrary short-range RBM state. For one thing, given a specific quantum many-body system, there is so far no systematic way to write down its wave function in terms of RBM. For another, we need certain symmetries (such as the translational symmetry) to substantially simplify the equations. Thus, this approach works only in certain specific circumstances.
%this approach only
%works for short-range RBM states with translational symmetry. For
%another, given a specific quantum many-body system, there is so far
%no systematic way to write down its wave function in terms of RBM.
However, we emphasize that our method does not contain sophisticated
mathematics and is a completely new approach never considered before in the literature. \\%It would inspire more
%studies along this direction and provide invaluable guidance for numerical
%simulations with machine learning techniques. 

\section{Conclusion and outlook}

In summary, we have studied the entanglement properties of neural-network
quantum states in the RBM architecture. In particular, we have proved
that all short-range RBM states satisfy an area law of entanglement
for arbitrary dimensions and bipartition geometry. This not only immediately
implies an area law for the entanglement of the 1D SPT cluster states and
the 2D/3D toric code states (with or without anyonic excitations), but also sheds  light on the open
problem of proving the entanglement area law for the ground states
of local gapped Hamiltonians in higher dimensions. For generic long-range
RBM states with random parameters, we numerically studied their entanglement
entropy and spectrum. We found: (i) the averaged entanglement entropy
follows a volume law, but is significantly smaller than the Page-entropy
for random pure states; (ii) their entanglement spectrum has \textit{no}
universal part associated with random matrix theory and manifests a Poisson-type
level statistics. In addition, we analytically constructed families
of RBM states (in both 1D and 2D) with maximal volume-law entanglement,
which cannot be represented efficiently in terms of matrix product states
or tensor-network states. For these states, the RBM representation
is remarkably efficient, requiring only a small number of parameters
 scaling \textit{linearly} with the system size. These results explicitly
show, in an \textit{exact} fashion, the remarkable power of artificial
neural networks in describing  quantum states with massive entanglement. Unlike MPS or tensor-network states, entanglement is not the limiting factor for the efficiency of the  neural-network representation of quantum many-body states.  Through reinforcement learning of a modified Haldane-Shastry model, we have shown that RBM is capable of calculating the ground state, which has power-law entanglement. The corresponding ground-state energy and correlations can also be efficiently obtained. Finally, we also demonstrated, through a concrete example, that the RBM representation could be
used as a tool to analytically compute the entanglement entropy and
spectrum for  finite systems. Our results reveal some crucial
aspects of the data structures of neural-network quantum states and
 provide a  useful guide for the practical applications
of machine learning techniques in solving quantum many-body problems.

There remain many open questions.  First, what are the limiting factors for RBMs in efficiently representing quantum states? In the future, it would be interesting and important to find out the necessary and sufficient conditions
under which a many-body state can be represented efficiently by neural networks
and find out how to convert a general quantum state satisfying these conditions into RBMs. This
would help develop new machine-learning algorithms for solving many-body
problems and advance the understanding, from a physical perspective
\cite{Lin2016Why}, of the power of machine learning itself.  It would also be interesting to study entanglement properties in other types of artificial-neural-network states \cite{Carrasquilla2016Machine,van2016Learning}. Another
interesting direction worth more investigations is the relation between
the MPS/tensor-network representation and the neural-network representation.
In this context, we note a recent work on supervised machine learning
with quantum-inspired tensor networks \cite{Stoudenmire2016Supervised}.
From Sec. III, we now know that all short-range RBM states in 1D can
be represented in terms of MPS. What about higher dimensions
and the inverse statement? Can we rewrite all states bearing a MPS/tensor-network representation with small bond dimensions in
terms of short-range RBMs? These are among the questions worth future exploration. 

\section{Acknowledgments}
We thank Frank Verstraete, Roger Melko, Lei Wang, Giuseppe Carleo, Zlatko Papic, Ignacio Cirac, Mikhail D. Lukin, Subir Sachdev, and Matthias Troyer for helpful discussions.  D.L.D. thanks Zhengyu Zhang for his help in developing the code for the reinforcement learning calculations.
This work is supported by JQI-NSF-PFC and LPS-MPO-CMTC. 
X.L. acknowledges support by the Start-Up Fund of Fudan University. D.L.D. thanks the hospitality of the Kavli Institute for Theoretical Physics (KITP), where this work was partially done.
We acknowledge
the University of Maryland supercomputing resources (http://www.it.umd.edu/hpcc)
made available in conducting the research reported in this paper.

\bibliographystyle{apsrev4-1-title}
\bibliography{/Users/dldeng/Documents/DLDENG/Dropbox/Deng-DataBasis/tex/Dengbib}

\end{document}